\documentclass[12pt]{article}
\usepackage{amsmath,amssymb,amsthm,amsfonts,bm,mathrsfs,float,algorithm,caption,subcaption,setspace}
\usepackage{avant,array,geometry,fontenc,inputenc,natbib,hyperref,multirow,enumerate,rotating,booktabs,color,latexsym,times,stmaryrd,xr}
\usepackage{mathtools}

\usepackage[]{algpseudocode}

\newtheorem{theorem}[]{Theorem}

\newtheorem{proposition}[]{Proposition}

\newtheorem{example}[]{Example}

\makeatletter
\newcommand*{\addFileDependency}[1]{
  \typeout{(#1)}
  \@addtofilelist{#1}
  \IfFileExists{#1}{}{\typeout{No file #1.}}
}
\makeatother

\newcommand*{\myexternaldocument}[1]{%
    \externaldocument{#1}%
    \addFileDependency{#1.tex}%
    \addFileDependency{#1.aux}%
}

\myexternaldocument{2-arxiv-supp}

\begin{document}

\title{\Large{\textbf{Multi-Layer Kernel Machines: Fast and Optimal Nonparametric Regression with Uncertainty Quantification}}} 
\author{
\bigskip
{\sc Xiaowu Dai$^{1}$} \\
{\it {\normalsize University of California, Los Angeles}}
\and
\bigskip
{\sc Huiying Zhong} \\
{\it {\normalsize Peking University}}
}
\date{}
\maketitle
\begin{footnotetext}[1]
{\textit{Address for correspondence:} Xiaowu Dai, Department of Statistics and Data Science and Department of Biostatistics, UCLA, 8917 Math Sciences Bldg \#951554,  Los Angeles, CA 90095,  USA. Email: dai@stat.ucla.edu.}
\end{footnotetext}

\begin{abstract}
Kernel ridge regression (KRR) is widely used for nonparametric regression over reproducing kernel Hilbert spaces. It offers powerful modeling capabilities at the cost of significant computational costs, which typically require $O(n^3)$ computational time and $O(n^2)$  storage space, with the sample size $n$. We introduce a novel framework of multi-layer kernel machines that approximate KRR by employing a multi-layer structure and random features, and study how the optimal number of random features and layer sizes can be chosen while still preserving the minimax optimality of the approximate KRR estimate. For various classes of random features, including those corresponding to Gaussian and Mat\'{e}rn kernels, we prove that multi-layer kernel machines can achieve $O(n^2\log^2n)$ computational time and $O(n\log^2n)$ storage space, and yield fast and minimax optimal approximations to the KRR estimate for nonparametric regression. Moreover, we construct uncertainty quantification for multi-layer kernel machines by using conformal prediction techniques with robust coverage properties. The analysis and theoretical predictions are supported by simulations and real data examples.
\end{abstract}
\bigskip


\newpage
\baselineskip=21pt

\section{Introduction}
Consider a collection of $n$ samples $\{(x_i,y_i)\}_{i=1}^n$ of covariate-response pairs, where the covariate vector $x_i\in \mathcal{X}\subset\mathbb{R}^d$ and the response variable $y_i\in \mathbb{R}$. In the standard nonparametric model, it is assumed that the covariate-response pairs follow the regression model,
\begin{equation}
\label{eqn:trueregression}
   y_i=f^*(x_i)+\epsilon_i,\quad i=1,\ldots,n,
\end{equation}
where the sequence $\{\epsilon_i\}_{i=1}^n$ consists of i.i.d. centered errors with $\mathbb E[\epsilon_i^2]=\sigma^2$. 
Our goal is to estimate the regression function $f^*$. It is typical to assume that $f^*$ has certain smoothness properties, such as belonging to a reproducing kernel Hilbert space (RKHS) \cite{aronszajn1950theory, wahba1990spline}. Given such assumptions, the \emph{kernel ridge regression} (KRR) method \cite{Hastie2008THE}  estimates $f^*$ in an RKHS $\mathcal H$ by minimizing a combination of the least-squares fit to the data and a penalty of the squared RKHS norm,
\begin{equation}\label{eqn:optrkhs}
\widehat{f}^{\text{KRR}} = \underset{f\in\mathcal H}{\arg\min}\left\{\frac{1}{n}\sum_{i=1}^n[y_i-f(x_i)]^2+\lambda\|f\|_{\mathcal H}^2\right\},
\end{equation} 
where $\lambda> 0$ is a tuning parameter and $\mathcal H$ corresponds to a positive definite kernel $K(\cdot,\cdot):\mathcal{X}\times \mathcal{X}\rightarrow \mathbb R$. 
By the representer theorem \cite{KIMELDORF197182}, the estimator \eqref{eqn:optrkhs} can be written as,
\begin{equation}\label{eq:krr}
    \widehat{f}^{\text{KRR}}(x)=\sum_{i=1}^n \alpha_i K(x,x_i),
\end{equation}
where $\alpha=(\mathbf{K}+\lambda n I)^{-1}Y$, $\mathbf{K}\in\mathbb R^{n\times n}$ is the kernel matrix with entries $K(x_i,x_j)$, and $i,j=1,\ldots,n$. The response vector $Y=(y_1,\cdots,y_n)^\top$. 

Based on existing results of $M$-estimation and empirical processes, it is known that when the parameter $\lambda$ in \eqref{eqn:optrkhs} is set appropriately, the KRR estimator $\widehat{f}$ achieves the minimax estimation rate for various classes of kernels \cite{geer2000empirical, mendelson2002geometric}. 
However, the computational complexity of computing KRR prevents it from being routinely 
used in large-scale applications that have large sample sizes $n$. 
The optimization in (\ref{eqn:optrkhs}) is an $n$-dimensional quadratic programming, which can be solved at  $O(n^3)$  time complexity and $O(n^2)$ space complexity through standard QR decomposition methods \citep{saunders1998ridge}. 
Therefore, it is important to develop methods for approximating the KRR estimate that still retain guarantees of statistical minimax optimality. 
Various methodologies have been developed to address this challenge, including random projections of kernel matrix \cite{halko2011finding, 2017yangrandomized} and random feature mappings \cite{2007Random, 2021Random}. Specifically, Rahimi and Rechet \cite{2007Random} propose a random feature mapping approach of KRR, in which the random Fourier features $\phi(x):\mathcal{X}\rightarrow \mathbb R^D$ approximate the reproducing kernel $K(x,x')=\mathbb E [\phi(x)^\top\phi(x')]$.
The KRR estimator \eqref{eq:krr} can be approximated by a standard ridge regression estimator,
\begin{equation}\label{eq:rff}
    \widehat{f}^{\text{RF}}(x)=c^\top \phi(x),
\end{equation}
where $c=(\mathbf{\Psi}^\top\mathbf{\Psi})^{-1}\mathbf{\Psi}^\top Y$, and the feature matrix $\mathbf{\Psi}=(\phi(x_1),\cdots,\phi(x_{n}))^\top\in\mathbb R^{n\times D}$. 
This random feature mapping approach has time complexity and space complexity $O(nD^2)$ and $O(nD)$, respectively. Rudi and Rosasco \cite{Rudi2016GeneralizationPO} give conditions on the number of samplings $D$, as a function of the kernel, under which minimax optimality of the random feature mapping estimator \eqref{eq:rff} can be guaranteed.

In general, classical random feature mapping works very well when the underlying problem fits the chosen reproducing kernel space. However, the resulting estimator may fail to approximate the true function otherwise. Then, if the appropriate reproducing kernel space is unknown a priori, one usually relies on so-called multiple kernel learning algorithms \cite{lanckriet2004learning, zhang2006component, bach2008consistency, koltchinskii2010sparsity}, which combine various features captured by different kernels $K_l$ with $l=1,\ldots,L$. While these methods facilitate learning a suitable kernel by constructing a linear or convex combination of input kernels, they still do not achieve significantly better results than standard a priori kernel choices for many applications  \cite{gonen2011multiple}.

In this paper, we introduce a new method called the Multi-Layer Kernel Machine (MLKM), which combines random feature mapping with multiple kernel learning into a multi-layer structure.
Our approach offers two main advantages over traditional multiple kernel learning:  first, it maps each kernel $K_l$ into a random feature space using $D_l$ random Fourier features to enhance computational efficiency; second, it aggregates these kernel estimates using a multi-layer structure and cross-fitting scheme to increase estimation accuracy. 
The MLKM process involves generating multi-scale random features, where the lower layer corresponds to a larger number of random features and the deeper layer to a smaller number. 
This approach regarding the number of random features is motivated by the concept of downsampling in the literature of wavelets \citep{daubechies1992ten, mallat2016understanding}.
Then, MLKM combines features layer by layer, divides data into disjoint subsamples, iteratively updates parameters using a cross-fitting scheme, and finally constructs the estimator through cross-fitting. We demonstrate the conditions under which the number $D_l$ of random features at each layer $l$  allows the approximate KRR estimate to retain the minimax optimality.
Consequently, we show that MLKM achieves the time and space complexity at $O(n^2\log^2n)$ and $O(n\log^2n)$, respectively, which are significantly better than those of classical random feature mapping and kernel ridge regression methods. 

We further extend MLKM to a Residual Kernel Machine (RKM) framework by employing residual learning techniques proposed by \cite{He2015DeepRL}. This integration addresses the challenges of vanishing or exploding gradients in training multi-layer kernels and allows for enhanced estimation accuracy.
Additionally, we study the uncertainty quantification for MLKM. Our approach is based on conformal prediction using weighted residuals, and it differs from existing inference methods for kernel machines \citep{wahba1983bayesian, Dai2021OrthogonalizedKD, zhou2023inference, zhou2024estimation}. We demonstrate that the constructed MLKM confidence band guarantees finite-sample marginal coverage without assuming specific noise structures. 

Our multi-layer kernel machines are different in some important ways from the classical deep kernel learning \cite{cho2009kernel, Wilson2015DeepKL, 2016Learning,bohn2017representer}. Classical deep kernel learning concatenates a kernel function with one or more nonlinear functions in order to construct a highly flexible new kernel function. Here the nonlinear function is typically a composition of functions represented by a deep neural network.
Our work differs substantially from deep kernel learning by: (i) targeting general, nonlinear regression problems through the application of random feature mappings of different kernels; (ii) establishing a minimax optimal bound on the expected risk. 
Additionally, our work is different from another class of methods that construct deep Gaussian processes \cite{damianou2013deep,2017How}. These methods recursively layer Gaussian processes to form deeper models, where the inference is based on approximate Bayesian inference techniques. Such inference often requires the optimization of a variational lower bound on the marginal likelihood, which process can become computationally intractable.
In contrast, our method employs a penalized likelihood estimation and leverages a conformal prediction framework for uncertainty quantification. Hence, our multi-layer kernel machines provide a practical and useful alternative to the classical deep kernel learning and deep Gaussian processes methods.

The rest of the paper is organized as follows. Section \ref{sec:2} first introduces the structure of the multi-layer kernel machine (MLKM), then develops the training algorithm. Section \ref{sec:convrates} presents the statistical optimality of MLKM by establishing rigorous convergence rates. Section \ref{sec:infer} illustrates the procedure to construct the confidence band and its finite-sample guarantee. We conduct simulations and real data analysis in Section \ref{sec:numerical} and Section \ref{sec:real} to validate the estimation and inference performance of our proposed MLKM. Finally, Section \ref{sec:summary} summarizes our work. Technical proofs and additional numerical examples are provided in the Supplementary Appendix.

\section{Multi-Layer Kernel Machines}\label{sec:2}
The construction of the multi-layer kernel machines (MLKM) consists of five main steps. Step 1 is to generate multiple scales of random features. Step 2 is to construct a multi-layer kernel machine to combine these random features layer by layer. Step 3 is to divide the data into $L$ disjoint subsamples. Step 4 involves the iterative parameter updates, where a new estimator for parameters in the $l$-th layer is obtained at each iteration. This step is carried out repeatedly over a number of epochs until a stopping criterion is met. Finally,  Step 5 is to construct the final estimator using cross-fitting. We summarize our procedure in Algorithm \ref{alg0} first, then discuss each step in detail.

\begin{algorithm}[ht]
\caption{Multi-layer kernel machines for multi-scale nonparametric regression}
\label{alg0}
\begin{algorithmic}[1]
\Require Training data $\{(x_i,y_i)\}_{i=1}^n$,  the number of layers $L$, and the number of epochs $T$ in stopping criterion.
\State Sample multi-scale random features $\phi_l(x)$ for $l=1,\cdots,L$ using \eqref{eqn:rfsigmal}.
\State Construct the MLKM \eqref{fun1} with the model parameters $\mathbf{W}=(W_1,\cdots,W_L)$.
Initialize $\mathbf{W}$ with a uniform distribution.  Reorganize the subsamples in rotational orders in \eqref{eqn:datasplit}.
\State  Split the data randomly into $L$ non-overlapping parts $\mathcal{I}_1,\cdots,\mathcal{I}_L$ with equal size. For $t\in \{1,\cdots,L\}$, denote $\mathcal{I}_{L+t}=\mathcal{I}_t$. Run the following algorithm of Alternating Direction Descent with Subsamples (ADDS).
\Repeat 
   \For{$j=1,\cdots,L$}
   \State Rotate the subsample as $(\mathcal{I}_{j_1},\cdots,\mathcal{I}_{j_L})$ where $j_l=j+l-1$.
       \For{$l=1$ to $L$}
        \State Obtain the estimator $\widehat{W}_l^{(j)}$ using the sample $\mathcal{I}_{j_l}$ through \eqref{loss-split} and \eqref{gd-split} while fixing other parameters $\{W_1^{(j)},\cdots,W_{l-1}^{(j)}, W_{l+1}^{(j)},\cdots,W_{L}^{(j)}\}$.
        \EndFor
   \EndFor
\Until{the overall loss \eqref{overall-loss} has no significant improvement over $T$ consecutive epochs.}
\Ensure  Construct the final estimator $\widehat{f}= \dfrac{1}{L} \sum_{j=1}^L \widehat{f}^{(l)}(x, \widehat{\mathbf{W}}^{(l)})$ in \eqref{fun3}.
\end{algorithmic}  
\end{algorithm}

\subsection{Multi-layer kernel machines}
We construct a multi-layer architecture of $L$ layers, which includes the input layer, the output layer, and $L$ hidden layer. Each layer $l$ is associated with a reproducing kernel $K_l$ with the scaling parameter chosen as $c_0\gamma^l$  for $l=1,\ldots,L$, where  $c_0>0$ and $\gamma>1$ are constants. We can choose various kernels to form the multi-layer structure. Examples include
\begin{equation}
\label{eqn:kerneleg}
\begin{aligned}
& \text{Gaussian kernel}\quad && K_l(x,x')= a_{1,l}\exp{(-\|x-x'\|^2/2(c_0\gamma^l)^2)},\\
&\text{Mat\'{e}rn kernel}\quad && K_{l,\nu}(x,x')=a_{2,l}\frac{2^{1-\nu}}{\Gamma(\nu)}\left( \frac{\sqrt{2\nu}\|x-x'\|}{c_0\gamma^l} \right)^\nu J_\nu \left( \frac{\sqrt{2\nu}\|x-x'\|}{c_0\gamma^l} \right).
\end{aligned}
\end{equation}
Here $l=1,\ldots,L$ and $a_{1,l},a_{2,l}$ are the normalization constants. For Mat\'{e}rn kernel, $\nu$ is a kernel parameter and $J_\nu$ is a modified Bessel function.

We approximate the reproducing kernel $K_l$ in each layer using random feature mapping \citep{2007Random, 2021Random, dai2022kernel}, which enables us to construct a projection operator between the RKHS and the original predictor space. Specifically, if the kernel functions are shift-invariant, i.e., $K_l(x,x')=K_l(x-x')$, and integrate to one, i.e., $\int_{\mathcal X}K_l(x-x')d(x-x')=1$, then the Bochner's theorem states that such kernel functions satisfy the Fourier expansion:
\begin{equation*}
K_l(x-x')=\int_{\mathbb R^d}p_l(\omega)e^{\sqrt{-1}\omega^{\top}(x-x')}d\omega,
\end{equation*}
where $p_l(w)$ is a probability density defined by
\begin{equation}
\label{eqn:defofpl}
p_l(\omega) = \int_{\mathcal X}K_l(x)e^{-2\pi\sqrt{-1}\omega x}dx.
\end{equation}
Note that many kernels are shift-invariant and integrate to one, including the examples in \eqref{eqn:kerneleg}.
The random features $\psi_k$s are constructed as follows. 
Let $\omega$ and $b$ be independently drawn from $p_l(\omega)$ and  $\text{Uniform}[0,2\pi]$, respectively, and  $z_\omega(x)=\sqrt{2}\cos(\omega^{\top} x+b)$. Then we can obtain $D$ randomly selected $z_{w_k}(x)$, $k=1,\cdots,D$, and let $\psi_k(x)=z_{\omega_k}(x)/\sqrt{D}$.

\begin{figure}[htbp]
	\centering
    \begin{minipage}[t]{0.36\textwidth}
		\centering
		\includegraphics[width=0.95\textwidth]{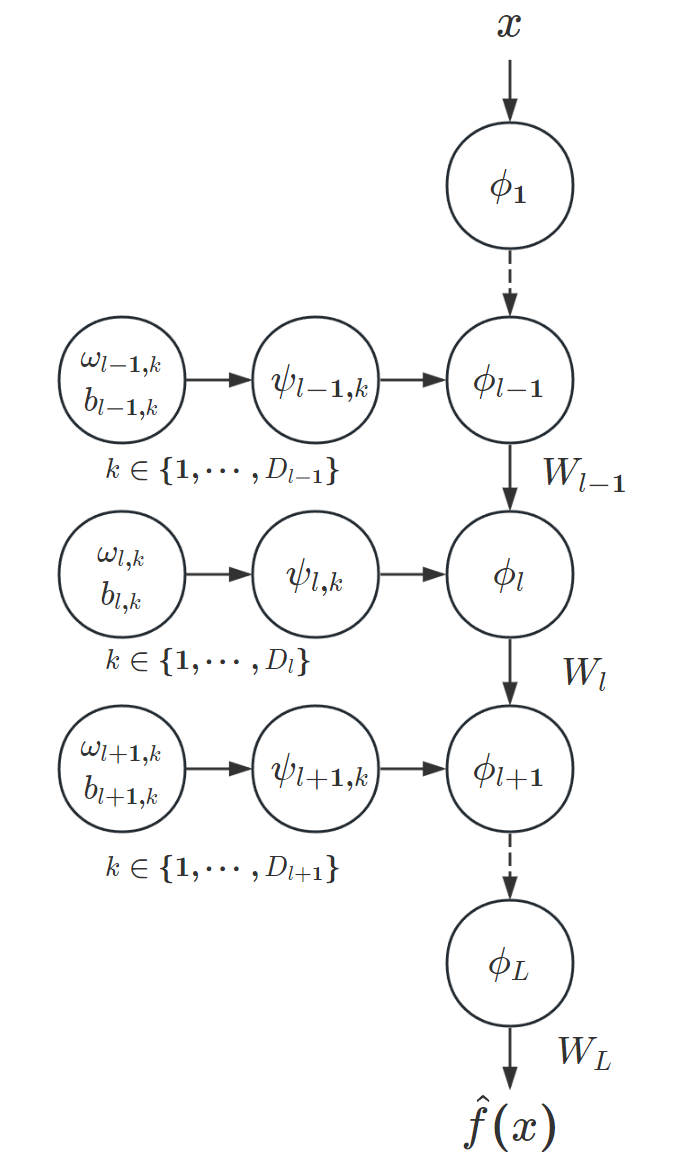}
        \subcaption{}
        \label{fig:structure(a)}
	\end{minipage}
    \begin{minipage}[t]{0.62\textwidth}
		\centering
		\includegraphics[width=0.95\textwidth]{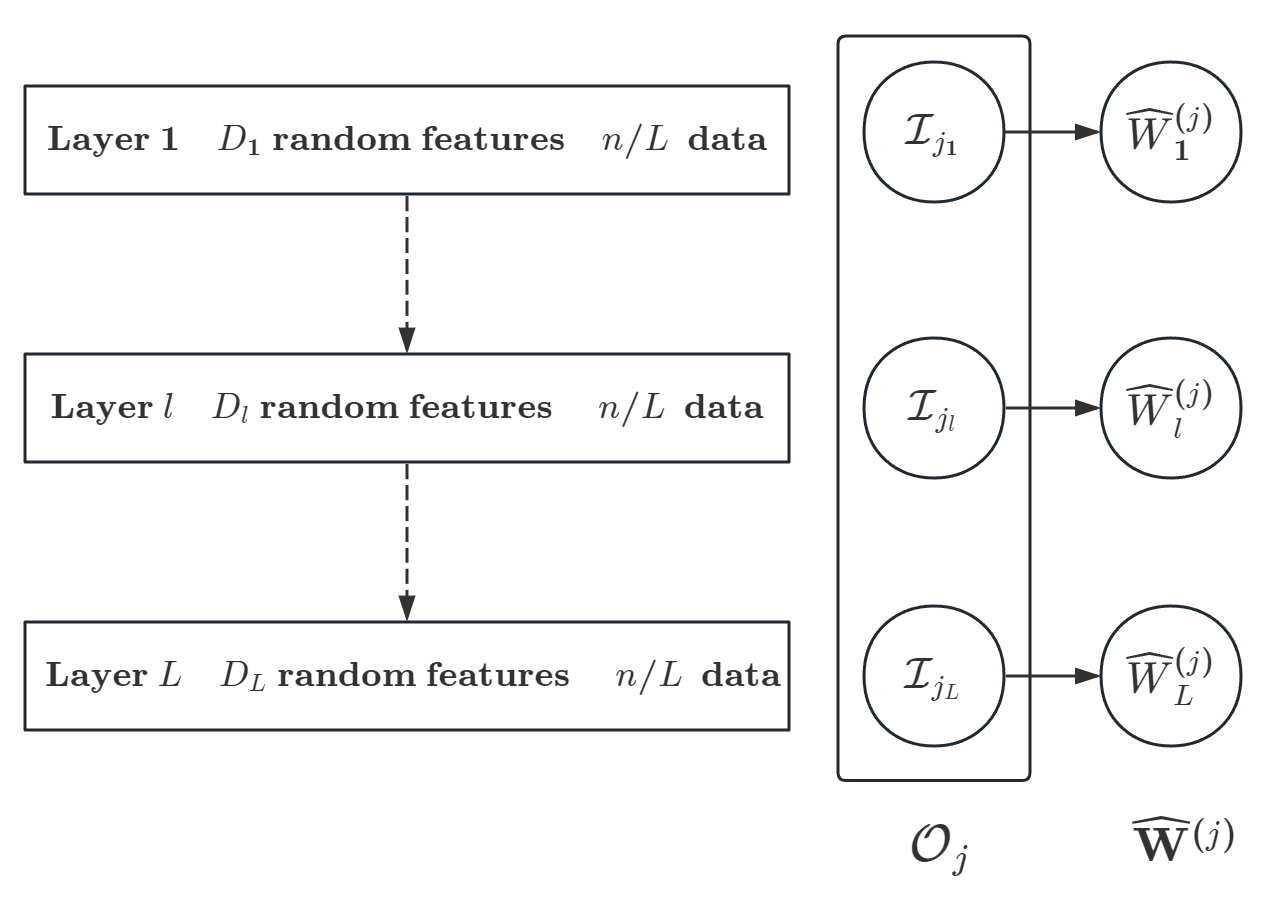}
        \subcaption{}
        \label{fig:structure(b)}
	\end{minipage}
	\caption{\small{Multi-Layer Kernel Machine (MLKM): Plot (a) shows the general architecture of MLKM that maps a $d$-dimensional inputs $x$ through a cascade of $L$ hidden layers parametrized by $\mathbf{W}=\{W_1,\cdots,W_L\}$. In each layer, we construct $D_l$ random features. Plot (b) illustrates the cross-fitting procedure where parameters in each layer are trained by a specific subsample of $n/L$ data. Using a specific subsample order $\mathcal{O}_j=(\mathcal{I}_{j_1},\cdots,\mathcal{I}_{j_L})$ yields an estimator $\widehat{\mathbf{W}}^{(j)}=\{\widehat{W}_1^{(j)},\cdots,\widehat{W}_L^{(j)}\}$.}}
	\label{fig:structure}
\end{figure}

We now construct the \emph{Multi-Layer Kernel Machine} (MLKM). For each layer $l$, we sample a set of $D_l$ random features $\{\psi_k(x),k=1,\cdots,D_l\}$ and let  
\begin{equation}
\label{eqn:rfsigmal}
\phi_l(x)=(\psi_1(x),\cdots,\psi_{D_l}(x))^\top,\quad  \forall l=1,\ldots,L.
\end{equation}  
The MLKM estimator is given by,
\begin{equation}\label{fun1}
    f(x)=W_L\phi_L \left[ W_{L-1}\phi_{L-1}(\cdots W_1\phi_1(x))\right].
\end{equation}
The matrix $W_l\in\mathbb R^{D_{l+1}\times D_{l}} $  represents the linear transformation, where $D_{l}>D_{l+1}$ for $l=1,\ldots,L$ and  $D_{L+1}=1$.
The MLKM approach involves introducing multiple layers, each encompassing different random features characterized by varying scales. The linear combination of random features in each layer is then propagated to the subsequent layers. 
In the rest of the paper, we use $\mathbf{W}=\{W_1,\cdots,W_L\}$ to denote the model parameters. The MLKM estimator can be equivalently written as $ f(x) =  f(x; \mathbf{W})$, where an illustration is shown in  Figure \ref{fig:structure(a)}.

\subsection{The ADDS algorithm}
\label{sec:ADDS}
We adopt the cross-fitting scheme similar to that in \cite{2017Double, Dai2021OrthogonalizedKD}. 
Specifically, we randomly split the sample into $L$ non-overlapping parts $\{\mathcal{I}_1,\cdots,\mathcal{I}_L\}$, where $\mathcal I_l$ has a sample size $n_l$ and $l=1,\ldots,L$. For simplicity, we take $n_l=n/L$ of equal split-sizes. 
We introduce $\mathcal{O}_j=(\mathcal{I}_{j_1},\cdots,\mathcal{I}_{j_L})$ that collects subsamples in rotational orders, where $j=1,\ldots,L$ and $j_l=j+l-1$. That is, 
\begin{equation}
\label{eqn:datasplit}
\mathcal{O}_1=(\mathcal{I}_1,\mathcal{I}_2,\cdots,\mathcal{I}_L), \ 
\mathcal{O}_2=(\mathcal{I}_2,\cdots,\mathcal{I}_L,\mathcal{I}_1), \ 
\ldots, \ 
\mathcal{O}_L=(\mathcal{I}_L,\mathcal{I}_1,\cdots,\mathcal{I}_{L-1}).
\end{equation}

We propose a new algorithm called the \emph{Alternating Direction Descent with Subsamples} (ADDS) to compute the MLKM estimator $f(x)$ in \eqref{fun1}. For each $\mathcal O_j$ with $j=1,\ldots,L$, we apply an iterative procedure to update the parameters in one layer while fixing the parameter values in other layers. Specifically, at the $l$th iteration ($l=1,\ldots,L$), we use the subsample  $\mathcal{I}_{j_l}$ to update the parameters at the $l$th layer using the gradient descent algorithm, where the loss function for regression problems is chosen as the mean squared error:
\begin{equation} \label{loss-split}
    \text{Loss}_l^{(j)}=\dfrac{1}{|\mathcal{I}_{j_l}|}\sum_{(x,y)\in \mathcal{I}_{j_l}}\left[y-f(x;W_1^{(j)},\cdots,W_{l-1}^{(j)},\widehat{W}_l^{(j)}, W_{l+1}^{(j)},\cdots,W_{L}^{(j)})\right]^2.
\end{equation}
Here, we keep $W_1^{(j)},\cdots,W_{l-1}^{(j)},W_{l+1}^{(j)},\cdots,W_{L}^{(j)}$ fixed and only update the parameter $\widehat{W}_l^{(j)}$. The gradient descent step is updated according to,
\begin{equation} \label{gd-split}
    \widehat{W}_{l,t+1}^{(j)} = \widehat{W}_{l,t}^{(j)}-\eta \dfrac{\partial \text{Loss}_l^{(j)}}{\partial W_l^{(j)}}, \quad t\geq 1,
\end{equation}
where $\eta>0$ is the learning rate. By updating the parameters from the layer $l=1$ to the layer $l=L$, we obtain an estimator 
\begin{equation*}
    \widehat{f}^{(j)}(x, \widehat{\mathbf{W}}^{(j)})=\widehat{W}_L^{(j)}\phi_L \left[ \widehat{W}_{L-1}^{(j)}\phi_{L-1}(\cdots \widehat{W}_1^{(j)}\phi_1(x))\right],
\end{equation*}
which is trained with sample $\mathcal O_j$. 
We perform the iterative process for 
$j=1,\cdots,L$ and continue iterating until the following stopping criterion is met. Figure \ref{fig:structure(b)} illustrates the cross-fitting procedure for each $\mathcal{O}_j$. Let the overall training loss be,
\begin{equation}\label{overall-loss}
    \text{Loss}^{\text{all}}=\dfrac{1}{nL}\sum_{j=1}^L\sum_{i=1}^n\left[y_i-\widehat{f}^{(j)}(x_i, \widehat{\mathbf{W}}^{(j)})\right]^2.
\end{equation}
The training process is stopped when the overall loss in \eqref{overall-loss} has no significant improvement over $T$ consecutive epochs, where $T$ is chosen as $50$ or $100$ epochs in our experiments.

Using the ADDS algorithm, we can obtain $L$ estimates $\widehat{f}^{(j)}(x, \widehat{\mathbf{W}}^{(j)})$ for $j=1,\cdots,L$.
The final cross-fitting MLKM estimator is formulated as,
\begin{equation}\label{fun3}
    \widehat{f}(x)= \dfrac{1}{L} \sum_{j=1}^L \widehat{f}^{(j)}(x, \widehat{\mathbf{W}}^{(j)}).
\end{equation}
This estimator rotates the subsamples to yield the $L$ estimators, thus leveraging the entire dataset across all layers to enhance efficiency.

\subsection{Bias reduction through cross-fitting}
The cross-fitting estimator $\widehat{f}$ in \eqref{fun3} can effectively reduce bias. 
To illustrate this, we consider a two-layer MLKM expressed as $\widehat{f}(x) = \widehat{W}_2\phi_2(\widehat{W}_1\phi_1(x))$.
Then the bias of $\hat{f}$ is,
\begin{equation*}
    \mathbb E \widehat{f}-f=
    \underbrace{\mathbb E\left[(\widehat{W}_2-W_2)\phi_2(\widehat{W}_1\phi_1(x))\right]}_{\text{Bias I}}+
    \underbrace{\mathbb E\left[W_2(\phi_2(\widehat{W}_1\phi_1(x))-\phi_2(W_1\phi_1(x)))\right]}_{\text{Bias II}}.
\end{equation*}
Without splitting data in \eqref{eqn:datasplit}, the term $\mathbb E[(\widehat{W}_2-W_2)\phi_2(\widehat{W}_1\phi_1(x))]$ does not vanish due to the correlation between $\widehat{W}_1$ and $\widehat{W}_2$. However, after deploying different subsamples for estimating $\widehat{W}_1$ and $\widehat{W}_2$, 
\begin{align*}
    \mathbb E\left[(\widehat{W}_2-W_2)\phi_2(\widehat{W}_1\phi_1(x))\right] & =\mathbb E[\widehat{W}_2-W_2]\mathbb E[\phi_2(\widehat{W}_1\phi_1(x))].
\end{align*}
The term, Bias I, vanishes if $\widehat{W}_2$ is unbiased. The term, Bias II, arises for any estimators of $W_1$, and vanishes when $\phi_2$ takes a linear form. Hence, the cross-fitting technique helps reduce the total bias.

\subsection{Extensions of the MLKM framework}\label{sec:extensionMLKM}
\begin{figure}[H]
	\centering
	\includegraphics[width=4in]{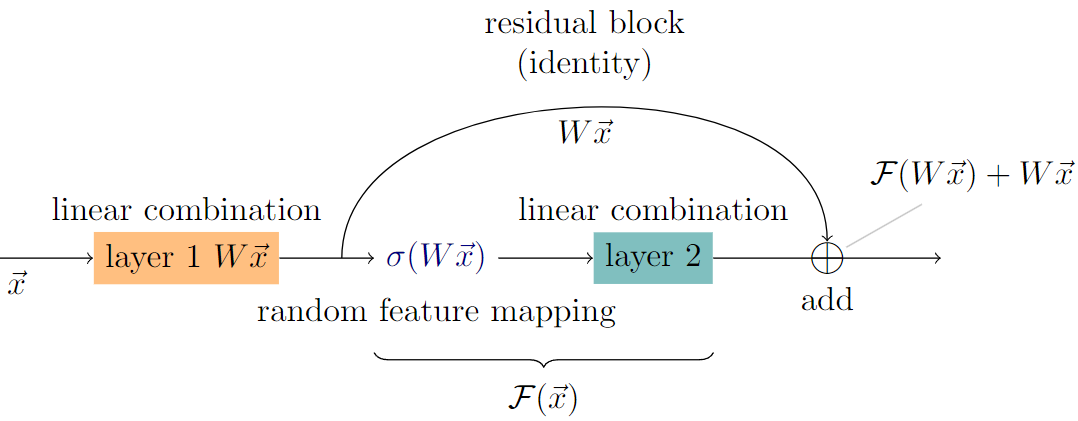}
	\caption{An illustration of the residual block in the residual-kernel machine.}
	\label{fig:residual}
\end{figure}

There are various approaches to extend the MLKM framework. 
First, we can incorporate the residual learning structure \citep{He2015DeepRL} into the proposed method.  
Specifically, we let for each $l=1,\ldots,L$,
\begin{equation*}
\mathcal{T}_l(z) = W_l^{(2)}\phi_l(W_l^{(1)}z)+W_l^{(1)}z \quad W_l=(W_l^{(1)},W_l^{(2)}).
\end{equation*}
Figure \ref{fig:residual} provides an illustration of the residual blocks in each layer. 
The \emph{Residual-Kernel Machine} (RKM) estimator is constructed as, 
\begin{equation}\label{fun2}
    \widehat{f}(x)=\mathcal{T}_L[\mathcal{T}_{L-1} \cdots (\mathcal{T}_2(\phi_1(x)))].
\end{equation}
We can apply the ADDS algorithm in Section \ref{sec:ADDS} to train the RKM estimator in \eqref{fun2}.
Compared to the MLKM estimator in \eqref{fun1}, the RKM estimator includes an additional residual block. 
We show in a numerical example of Section \ref{sec:numerical} that RKM can yield accurate and stable estimations.

Second, we can incorporate the regularization into MLKM. Take the $\ell_2$-regularization as an example. Instead of using \eqref{loss-split} directly, we consider an extra penalty term in the loss,
\begin{equation*}\label{l2_pen}
    \text{Loss}_l^{(j)}(\lambda)=\text{Loss}_l^{(j)}+\lambda \|\widehat{\mathbf{W}}^{(j)}\|_2^2,
\end{equation*}
where the tuning parameter $\lambda\geq 0$.
We can still apply the ADDS algorithm in Section \ref{sec:ADDS} to train our regularized models. The incorporation of regularization techniques and sparsity patterns avoids overfitting and may yield accuracy improvements \cite{Gale2019TheSO, SchmidtHieber2017NonparametricRU}. 

Finally, we recommend a layer-specific approach when selecting the scale parameters $\gamma$'s for random features across various layers. For the initial layers, we recommend choosing smaller $\gamma$ values and a larger number of random features $D_l$. 
For the deeper layers, it would be advantageous to choose larger $\gamma$ values and decrease the number of random features $D_l$. 
This approach of choosing $\gamma$ and the number of random features is motivated by the concept of downsampling in the literature of wavelets \citep{daubechies1992ten, mallat2016understanding}.

\section{Statistical Estimation}
\label{sec:convrates}
We study the estimation properties of the proposed MLKM method. In particular, we prove that MLKM achieves the optimal rates of convergence for nonparametric regressions. 

\subsection{Definitions}
Let $\rho:\mathcal{X} \times \mathbb{R} \rightarrow\mathbb{R}$ be the distribution function of $(X,Y)$ that has $n$ i.i.d. copies $\{(x_i,y_i)\}_{i=1}^n$. Let $\rho_X$ as the marginal distribution of the design $X$, and $\rho(\cdot|x)$ be the conditional distribution given $X=x$. 
Let $\mathcal{H}_l$ be the RKHS corresponding to the kernel function $K_l$, where $l=1,\cdots,L$. Each $\mathcal{H}_l$ has a bounded compact domain $\mathfrak{D}_l$ and range $\mathfrak{R}_l$. Here $\mathfrak{D}_1=\mathcal{X}$, $\mathfrak{R}_l\subseteq \mathfrak{D}_{l+1}\subseteq \mathbb R^{d_l}$, and $\mathfrak{R}_L\subseteq \mathbb R$. Define the function space $\mathcal{H}^{\text{comp}}$ as the RKHS composition, 
\begin{equation}\label{defH}
    \mathcal{H}^{\text{comp}}=\left\{f_L\circ f_{L-1}\cdots\circ f_1|f_l\in \mathcal{H}_l,l=1,\cdots,L\right\}.
\end{equation}
Define the population loss as, 
$\mathcal{L}(f)\equiv\mathbb{E}_{\rho}[f(X)-Y]^2$.
Assume that the true function $f^*$ in \eqref{eqn:trueregression} lies in the space $\mathcal{H}^{\text{comp}}$ in \eqref{defH}. Then 
\begin{equation*}
f^*=f_{\mathcal{H}^{\text{comp}}}=\underset{f\in\mathcal{H}^{\text{comp}}}{\arg\min}\ \mathcal{L}(f).
\end{equation*}
Define the excess risk as,
\begin{equation}
\label{eqn:excessrisk}
\mathcal{R}(f)\equiv\mathcal{L}(f)-\mathcal{L}(f_{\mathcal{H}^{\text{comp}}}).
\end{equation} 
The MLKM method in \eqref{fun1} considers the following compositional space for approximation,
\begin{equation}\label{defM}
    \mathcal{M}=\left\{f:f(x)=W_L\phi_L \left[ W_{L-1}\phi_{L-1}(\cdots W_1\phi_1(x))\right]\right\},
\end{equation}
where $\phi_l$ is the random feature mapping in \eqref{eqn:rfsigmal}. 
Then the best estimator in the space $\mathcal{M}$ is, 
$f_\mathcal{M}={\arg\min}_{f\in\mathcal{M}} \mathcal{L}(f)$.
Define the empirical loss as, $\mathcal{L}_{\mathcal{S}}(f)\equiv\sum_{i=1}^n [f(x_i)-y_i]^2/n$. Then, the optimal MLKM estimator is,
\begin{equation*}
\widehat{f}=\underset{f\in\mathcal{M}}{\arg\min}\ \mathcal{L}_{\mathcal{S}}(f). 
\end{equation*}


Recall that a function $f$ has H\"older smoothness index $q$ if: (1) when $q\leq 1$,  $f$ is $q$ H\"older smooth in the sense of existing a constant $C$ such that $|f(x)-f(y)|\leq C \|x-y\|_{\infty}^q$; (2) when $q>1$,  all partial derivatives of $f$ up to order $\lfloor q\rfloor$ exist and are bounded, and the partial derivatives of order $\lfloor q\rfloor$ are $q-\lfloor q\rfloor$ H\"older smooth. 
The concept of H\"older smoothness is closely related to the size of the reproducing kernel space \cite{Rudi2016GeneralizationPO}. A larger index $q$ corresponds to a benign setting with a smaller function class. 
For example, the kernels examples in \eqref{eqn:kerneleg} and beyond, including Gaussian, Mat\'{e}rn (with parameter $\nu$), Laplacian, and Cauchy kernels have H\"older smoothness indexes of $q=\infty, q=\nu, q=1/2, q=1$, respectively.
See more examples of H\"older smoothness in \citep{shekhar2020multi,rasmussen2006gaussian}.

\subsection{Convergence rates of estimation}
We now present our main result on the bounds of the excess risk. Given the multi-scale and multi-layered structure of the MLKM estimator, the bounds we derive rely on multiple scale parameters, which are different from those established in the literature \cite{Rudi2016GeneralizationPO}.

\begin{theorem} \label{excess}
Assume that density functions $p_l(\omega)$ in \eqref{eqn:defofpl} are sub-Gaussian for $l=1,\ldots,L$. Let $\mathcal{M}$ and $\mathcal{H}$ be function spaces defined in \eqref{defH} and \eqref{defM}. Assume that for any function $f_l \in \mathcal{H}_l$,  $\|f_l\|_{\mathcal{H}_l}\leq B$. For the weight vector $W_l$, it holds that $\|W_l\|_{1}\leq B$ and $\|W_l\|_{\infty}\leq B$.  
The loss function is bounded such that $[Y-f(X)]^2\leq \kappa<\infty$ for any $(X,Y)$ drawn from $\rho$ and $f\in \mathcal{M}$. Let $K_l$ be a kernel of $q_l$th H\"older smoothness. Define that
\[
\Delta_n = \min_{l=1,\cdots,L} \left( \frac{2q_l}{2q_l+d_l} \right) \prod_{t=l+1}^L (q_t \land 1).
\]
If the number of random features is given by 
\[
D_l \simeq n^{\frac{2q_l}{2q_l+d_l}} \log n,
\]
then we can bound the excess risk in \eqref{eqn:excessrisk} as 
\[
\mathcal{R}(\widehat{f}) \leq O_p(n^{-\Delta_n}).
\]
\end{theorem}
\noindent
We make three remarks on this theorem. Firstly, the assumption of bounded loss where $[Y-f(X)]^2\leq \kappa<\infty$ is common in the literature \cite[see,][]{yin2019rademacher}. 
Secondly, our result agrees with known optimal convergence rates in certain cases. For example, with a single-layer random feature method where $L=1$, Theorem \ref{excess} shows that the excess risk is $O_p(n^{-2q/(2q+d)})$. This rate is the same as the minimax optimal rate for this model \cite[see,][]{Rudi2016GeneralizationPO}.
On the other hand, Theorem \ref{excess} gives convergence rates that apply more broadly and incorporate different scale parameters, which capture different kernel smoothness in the multi-layer structure $L\geq 1$. 
Finally, we provide an outline of the proof.
The excess risk in \eqref{eqn:excessrisk} can be decomposed as,
\begin{equation*}
    \mathcal{R}(\widehat{f})=\underbrace{[\mathcal{L}(\widehat{f})-\mathcal{L}(f_{\mathcal{M}})]}_{\text{generalization error}} + \underbrace{[\mathcal{L}(f_{\mathcal{M}})-\mathcal{L}(f_{\mathcal{H}^{\text{comp}}})]}_{\text{approximation error}}.
\end{equation*}
For the generalization error, we use the McDiarmid Theorem \citep{vershynin2018high} to derive $ \mathcal{L}(\widehat{f}) - \mathcal{L}(f_{\mathcal{M}}) \leq O_p(n^{-\Delta_n})$. 
For the approximation error, we use the generalization properties in \cite{Rudi2016GeneralizationPO} and choose $D_l \simeq n^{2q_l/(2q_l+d_l)} \log n$ for all $l$ to obtain that $\sum_{i=1}^n[f_\mathcal{M}(x_i)-f_{\mathcal{H}^{\text{comp}}}(x_i)]^2/n\leq O_p(n^{-\Delta_n})$. By the empirical process theory and localized Rademacher complexity \cite{raskutti2014early, 2019Early}, we can bound that $ \mathcal{L}(f_\mathcal{M}) - \mathcal{L}(f_{\mathcal{H}^{\text{comp}}})\leq O_p(n^{-\Delta_n})$.
Combining these two errors, we derive the result in the theorem.

\subsection{Convergence of the optimization phase}
Consider the optimization phase of Algorithm \ref{alg0}. Note that without cross-fitting, it is a stochastic gradient descent (SGD) algorithm that updates parameters in all layers altogether. The update can be written as, 
\begin{equation}
\label{eqn:mlkmsgd}
\mathcal{L}_{\mathcal{S}}^\lambda(\widehat{f}_t)=\mathcal{L}_{\mathcal{S}}^\lambda(\widehat{\mathbf{W}}_t)=\mathcal{L}_{\mathcal{S}}+\lambda \|\mathbf{W}\|_2^2,\quad \widehat{\mathbf{W}}_{t+1}=\widehat{\mathbf{W}}_{t}-\eta \dfrac{\partial \mathcal{L}_{\mathcal{S}}^\lambda}{\partial \widehat{\mathbf{W}}}, \quad t\geq 1.
\end{equation}
where $\widehat{f}_t=\widehat{f}(x, \widehat{\mathbf{W}}_t)=\widehat{W}_{L,t}\phi_L [ \widehat{W}_{L-1,t}\phi_{L-1}(\cdots \widehat{W}_{1,t}\phi_1(x))]$. It is known that under certain conditions, SGD can converge for both convex and non-convex optimization problems \cite{bottou2018optimization}. Given these results, we demonstrate the convergence of \eqref{eqn:mlkmsgd} in the following theorem.

\begin{theorem} \label{nonconvex0}
    Let $\widehat{\mathbf{W}}_t$ be the parameter updates at the $t$th iteration in \eqref{eqn:mlkmsgd}. Under the same conditions of Theorem \ref{excess}, it holds that for any $\epsilon>0$,
    \begin{equation*}
    \dfrac{1}{T}\sum_{t=0}^{T-1}\|\mathcal{L}_{\mathcal{S}}^\lambda(\widehat{f}_t)\|_2^2=\dfrac{1}{T}\sum_{t=0}^{T-1}\|\nabla \mathcal{L}_\mathcal{S}^\lambda(\widehat{\mathbf{W}}_t)\|_2^2 \leq\epsilon^2,\quad \forall \lambda \geq 0,
    \end{equation*}
    where the step size $\eta$ is appropriately chosen, and the number of iterations $T=O(\epsilon^{-2})$.  
\end{theorem}

\subsection{Computational complexities}
\label{sec:compcomplexity}
We compare the computational complexity of MLKM with the kernel ridge regression (KRR) and random feature methods, where the results are summarized in Table \ref{table:compwithsingle}. We discuss each result in detail.

\begin{table}[H]
    \caption{Comparison of computational complexities, where the results are up to constants. For brevity, we denote $D=O(n^{2q/(2q+d)}\log{n})$, and $D_l=O(n^{2q_l/(2q_l+d_l)}\log{n}), \forall l$.}
\centering
    \begin{tabular}{c|c|c|c}
    \hline
     & KRR & Random Feature Mapping & MLKM \\
    \hline
    Storage Space  & $O(n^2)$  & $O(nD)$ & $O(\sum_{l=1,\ldots,L-1} D_lD_{l+1})$  \\
    \hline
    Computational Time & $O(n^3)$ &  $O(nD^2)$ & $O(n\sum_{l=1,\ldots,L-1} D_lD_{l+1})$ \\
    \hline
    \end{tabular}
\label{table:compwithsingle}
\end{table}

\begin{itemize}
\item KRR estimator in \eqref{eq:krr} requires the computation of $\alpha=(\mathbf{K}+\lambda n I)^{-1}Y$. It requires the $O(n^2)$space to store the kernel matrix $\mathbf{K}\in\mathbb R^{n\times n}$, and the $O(n^3)$ computational time to compute the inversion operation  \cite{wahba1990spline}.
\item Random feature mapping estimator is given by $ \widehat{f}(x)=\sum_{k=1}^D c_k \psi_k(x)$  \citep{2007Random}, which approximates the KRR estimator in \eqref{eq:krr}. Here $c_k$s are unknown coefficients, and $D$ is the number of random features. The computation complexity becomes $O(nD)$ in space and $O(nD^2)$ in time.  Moreover, it requires  $D= O(n^{2q/(2q+d)}\log{n})$ for achieving the optimal rate of convergence \cite{Rudi2016GeneralizationPO}.
\item MLKM estimator in \eqref{fun1} requires the storage of $\mathbf{W}=\{W_1,\cdots,W_L\}$, where each of matrix $W_l\in \mathbb R^{D_{l+1}\times D_l}$ and $D_l=O(n^{2q_l/(2q_l+d_l)}\log{n})$.  the construction of a random feature mapping $\phi_l$ in each layer, where $\phi_l\in\mathbb R^{D_l}$.
Thus, it requires the storage space $O(\sum_{l=1}^{L-1} D_lD_{l+1})$. Regarding the computational time, each iteration involves the calculation of the gradient in \eqref{loss-split}, taking $O(n\sum_{l=1}^{L-1}D_{l+1} D_l)$ in time. 
\end{itemize}
\noindent
We make three remarks on these complexity results. Firstly, we compare MLKM with KRR estimators. Observe that  
\begin{equation*}
D_l/n=O\left(n^{-d_l/(2q_l+d_l)}\log{n}\right)\rightarrow 0, \text{ as }n\rightarrow\infty,
\end{equation*}
and with a fixed $L$, 
$\sum_{l=1}^{L-1} D_{l+1}D_l/n^2\rightarrow 0$.
Hence, MLKM provides substantial computational benefits over KRR estimator in both storage and computational time.

Secondly, to further demonstrate the computational advantages of MLKM over KRR, consider the following example.  For a kernel $K_l:\mathbb R^{d_l}\times\mathbb R^{d_l}\to\mathbb R$ with $q_l$th H\"older smoothness, the decay of its $i$th eigenvalue is characterized by $O(i^{-2q_l/d_l})$, where $0\leq d_l/(2q_l)\leq 1$. A fast decay (i.e., $d_l/(2q_l)\to 0$) corresponds to a smaller RKHS, whereas a slow decay (i.e., $d_l/(2q_l)\to 1$) indicates a larger RKHS \cite{wahba1990spline}. We have the following proposition by Table \ref{table:compwithsingle}.
\begin{proposition}
\label{prop:complexityMLKM}
Consider a scenario where $d_l/(2q_l) = 1$ for any $l=1,\ldots,L$, MLKM only requires  $O(n\log^2n)$ in storage and $O(n^2\log^2n)$ in computational time.
\end{proposition}
\noindent
The complexity result in Proposition \ref{prop:complexityMLKM} shows that MLKM
is significantly more efficient than KRR, which requires $O(n^2)$ storage and $O(n^3)$ computational time.

Finally, we compare MLKM with random feature mapping estimators.
Note that 
\begin{equation*}
\begin{aligned}
& D_{l+1}D_l/(nD)=O\left(n^{-1-2q/(2q+d)+2q_l/(2q_l+d_l)+2q_{l+1}/(2q_{l+1}+d_{l+1})}\log{n}\right),\text{ and}\\ &nD_{l+1}D_l/(nD^2)=O\left(n^{-4q/(2q+d)+2q_l/(2q_l+d_l)+2q_{l+1}/(2q_{l+1}+d_{l+1}}\right).
\end{aligned}
\end{equation*}
Assuming $d_l/q_{l}=d/q$ for $l=1,\ldots, L-1$, it follows that
$\sum_{l=1}^{L-1} D_{l+1}D_l/(nD)\rightarrow 0$, and $\sum_{l=1}^{L-1} nD_{l+1}D_l/(nD^2) = O(1)$.
Thus, MLKM achieves considerable savings in storage while preserving computational time efficiency compared to the random feature method.

\section{Statistical Inference}\label{sec:infer}
We employ the conformal prediction \cite{vovk2005algorithmic, lei2018distribution, 2023A, barber2023conformal} for the inference of the proposed MLKM method. Specifically, for a new feature value $x_{n+1}$, we consider the construction of a prediction interval for $y_{n+1}$, where $(x_{n+1},y_{n+1})$ is an independent draw from $\rho$. Equivalent, given a nominal level $\alpha\in(0,1)$, we are interested in constructing a prediction band $C\subseteq \mathcal X\times \mathbb R$ based on the i.i.d. samples $\{(x_i,y_i)\}_{i=1}^n$such that,
\begin{equation}
\label{eqn:predband}
    \mathbb P(y_{n+1}\in C(x_{n+1}))\geq 1-\alpha,
\end{equation}
where for a point $x\in\mathcal X$ we denote $C(x) = \{y\in\mathbb R: (x,y)\in C\}$. The goal is to construct prediction bands as in \eqref{eqn:predband} for the MLKM estimator in \eqref{fun1}.

\subsection{Conformal MLKM prediction}
We follow the split conformal prediction framework in \cite{lei2018distribution,2023A}. 
The process begins with a random split of the set $\{1,\cdots,n\}$ into two subsets: $\mathcal{A}_1$ and $\mathcal{A}_2$, where $|\mathcal{A}_1|=n'=n-m$ and $|\mathcal{A}_2|=m$. 
Using $\{(x_i,y_i):i\in\mathcal{A}_1\}$, we construct a MLKM estimator $\widehat{f}$. Subsequently, we employ $\{(x_i,y_i): i\in\mathcal{A}_2\}$ to construct the prediction band $C(x_{n+1})$ for a new  $x_{n+1}$ as follows. Define that
\begin{equation}
\label{eqn:weightedresiduals}
    R_{i} = \frac{|y_i - \widehat{f}(x_i)|}{\widehat{\sigma}_y(x_i)}, \ i \in\mathcal{A}_2,\quad \text{and} \quad R_{n+1} = \frac{|y-\widehat{f}(x_{n+1})|}{\widehat{\sigma}_y(x_{n+1})}.
\end{equation}
Here $\widehat{\sigma}_y^2(x)$ is a variance estimator of $y$ for any sample $(x,y)$ drawn from distribution  $\rho$. The construct of  $\widehat{\sigma}_y^2(x)$ will be discussed in Section \ref{sec:varmlkm}.
We rank the weighted residuals $R_{i}$ for any $i\in\mathcal{A}_2$, and compute $\widehat{v}$ as the $\lceil (1-\alpha)(m+1)\rceil$-th smallest value in $\{R_i: i\in\mathcal{A}_2\}$. Formally, we let
\begin{equation*}
    \widehat{v}=\min \Big\{v:\sum_{i\in\mathcal{A}_2}{\mathbf 1}\{R_{i}\leq v\}=\lceil (1-\alpha)(m+1)\rceil\Big\}.
\end{equation*} 
Thus the conformal MLKM prediction interval at $x_{n+1}$ is,
\begin{equation}
\label{eqn:cMLKMpi}
    C(x_{n+1}) = \left\{y\in\mathbb R: R_{n+1} \leq \widehat{v} \right\}.
\end{equation}
The residual $R_{i}$ in \eqref{eqn:weightedresiduals} accounts for the heterogeneity present among data points.

\subsection{Variance estimator for MLKM}
\label{sec:varmlkm}
We construct a variance estimator $\widehat{\sigma}_y^2(x)$ used in \eqref{eqn:weightedresiduals}.
We begin by considering the single-layer kernel machines, where $L=1$ and $f(x) =  W_1\phi_1(x)$ with $W_1\in\mathbb R^{1\times D_1}$. 
Without loss of generality, let $\mathcal A_1 = \{1,\ldots,n'\}$.
Let $Y = (y_1,\ldots,y_{n'})^\top$ represent the response vector and $\mathbf{\Psi}=(\phi_1(x_1),\cdots,\phi_1(x_{n'}))^\top\in\mathbb R^{n'\times D_1}$ denote the design matrix. Following Algorithm \ref{alg0}, an unbiased estimator of $W_1^\top$ is obtained as $\widehat{W}_1^\top=(\mathbf{\Psi}^\top\mathbf{\Psi})^{-1}\mathbf{\Psi}^\top Y$, with variance 
$\text{Var}(\widehat{W}_1^\top)=\sigma^2(\mathbf{\Psi}^\top\mathbf{\Psi})^{-1}$.
For the sample $(x_{i},y_{i})$ drawn from $\rho$ with $i\not\in\mathcal A_1$, where $y_{i} = W_1\phi_1(x_{i})+\epsilon_{i}$, the variance estimator of $y_{i}$ can be given by, 
\begin{equation*}
\widehat{\sigma}_y^2(x_{i})=\widehat{\sigma}^2\left\{\phi_1(x_{i})^\top (\mathbf{\Psi}^\top\mathbf{\Psi})^{-1}\phi_1(x_{i})+1\right\}, \quad \forall i\not\in\mathcal A_1,
\end{equation*}
where $\widehat{\sigma}^2=\|Y-\mathbf{\Psi}\hat{W}_1^\top\|^2/(n'-D_1)$ is an estimator for $\sigma^2$ \cite[e.g.,][]{2012Linear}.

Next, we consider the variance estimator for the multi-layer kernel machines, where $L>1$ and $f(x)=f(x;\mathbf{W})$ with $\mathbf{W}=\{W_1,\cdots,W_L\}$ collecting $p$ model parameters. Denote $\mathbf{f}(\mathbf x,\mathbf{W})=(f(x_1,\mathbf{W}),\cdots,f(x_{n'},\mathbf{W}))^{\top}$. 
An estimator $\widehat{\mathbf{W}}$ of $\mathbf{W}$ can be obtained by Algorithm \ref{alg0} with sample $\{(x_i,y_i):i\in\mathcal A_1\}$.
By the linear approximation method \cite{1996Confidence}, we have the first-order approximation $f(x;\widehat{\mathbf{W}})=f(x;\mathbf{W}^*)+\nabla_{\mathbf W} f(x;\widehat{\mathbf{W}})^{\top}(\widehat{\mathbf{W}}-\mathbf{W}^*)$. 
For a new data point $(x_{i},y_{i})$ sampled from $\rho$ for $\forall i\not\in\mathcal A_1$ with $y_{i} = f(x_{i};\mathbf{W}^*)+\epsilon_{i}$, the variance estimator for $y_{i}$ can be given by, 
\begin{equation*}
    \widehat{\sigma}_y^2(x_{i})=\widehat{\sigma}^2\left\{\nabla_{\mathbf W} f(x_{i};\widehat{\mathbf{W}})^{\top}(\mathbf{F}^{\top}\mathbf{F})^{-1}\nabla_{\mathbf W} f(x_{i};\widehat{\mathbf{W}}) +1\right\},\quad\forall i\not\in\mathcal A_1
\end{equation*}
where $\widehat{\sigma}^2=\|Y- \mathbf f(\mathbf x;\widehat{\mathbf{W}})\|^2/(n'-p)$ is an estimator for $\sigma^2$,  and $\mathbf{F}=(\partial_{W_k} f(x_i;\widehat{\mathbf{W}}))_{ik}\in\mathbb R^{n'\times p}$.

\subsection{Finite-sample guarantee}
We show that the conformal MLKM prediction interval in \eqref{eqn:cMLKMpi} is guaranteed to deliver finite-sample marginal coverage without any assumption on the noises $\epsilon_i$'s in \eqref{eqn:trueregression}.

\begin{theorem}\label{th:confor}
      Suppose $\{(x_i,y_i)\}_{i=1}^n$ and $(x_{n+1},y_{n+1})$ are i.i.d., and the weighted residuals $\{R_{i}: i\in \mathcal{A}_2\}$ in \eqref{eqn:weightedresiduals} have a continuous joint distribution. Then the conformal MLKM prediction interval in \eqref{eqn:cMLKMpi} satisfies that,
\begin{equation*}
    1-\alpha \leq \mathbb{P}(y_{n+1}\in C(x_{n+1})) \leq 1-\alpha+\dfrac{1}{m+1}.
\end{equation*}
\end{theorem}
\noindent
Note that weighted residuals $R_i$ in \eqref{eqn:cMLKMpi} differ from those in \cite{lei2018distribution} and \cite{2023A}. 
Specifically, \cite{lei2018distribution} employs the estimated mean absolute deviation for weights, in contrast to our use of the estimated standard deviation in \eqref{eqn:cMLKMpi}. Meanwhile, \cite{2023A} also uses the estimated standard deviation but differs in its estimation approach that involves maximizing the likelihood of the conditional variance $\text{Var}(y|x)$. Our approach for estimating standard deviation, detailed in Section \ref{sec:varmlkm}, is easy to implement. We compute the standard deviation explicitly for MLKM estimators without an additional step of model fitting.

\section{Simulation Examples} \label{sec:numerical}
We present three numerical examples to study the finite sample properties of the proposed MLKM method. We also compare it with alternative methods, including kernel ridge regression, random feature method, and neural network method. 

\begin{example}
\label{eg:additive1}
We consider a multivariate additive model $f^*(x_i)=\sum_{j=1}^4f_j(x_{ij})$, where
\begin{align*}
    &f_1(x_1)=6\left[0.1\sin(2\pi x_1)+0.2\cos(2\pi x_1)+0.3\sin(2\pi x_1)^2+0.4\cos(2\pi x_1)^3+0.5\sin(2\pi x_1)^3\right],\\
    &f_2(x_2)=3(2x_2-1)^2, \quad f_3(x_3)=5x_3, \quad f_4(x_4)=4\sin(2\pi x_4)/(2-\sin(2\pi x_4)).
\end{align*}
This model is considered by \cite{zhang2006component, Lu2019Kernel}. 
Let $E_1,\cdots,E_d$ and $U$ follow i.i.d. Uniform$[0,1]$ and $X_j=(E_j+t U)/(1+t)$ for $j=1,\cdots,d$. 
We generate data $x_{1j},\ldots,x_{nj}$ that are i.i.d. copies of $X_j$.
The correlation between $X_j$ and $X_{j'}$ is $t^2/(1+t^2)$ for $j\neq j'$. We set $t=1$. The noise $\epsilon_i$ in \eqref{eqn:trueregression} follows $N(0,1)$. Let the dimension  $d\in\{4,8,16,32,64,128\}$ and the sample size $n=4000$.
We compare four models: (i) kernel ridge regression (KRR) \cite{wahba1990spline}, (ii) random feature (RF) method \cite{2007Random}, (iii) the proposed MLKM in Algorithm \ref{alg0}, and (iv) the proposed residual-kernel machine (RKM) in \eqref{fun2}. We use five-fold cross-validation to select the penalty parameters $\lambda$ for these models. These models share the same Gaussian kernel scale parameters $\gamma=1$. For RF, we select $D=500$ Gaussian features. For the MLKM and RKM methods, we use a 4-32-8-1 layer structure when $d=4$, an 8-64-8-1 layer structure when $d=8$, and a $d$-256-16-1 layer structure when $d\in\{16,32,64,128\}$. We randomly split the data into two subsets with sizes $\{n',m\}$, where $n'=m=2000$. 
We use $n'$ data to construct the MLKM machine and use $m$ data to construct the conformal MLKM prediction interval in \eqref{eqn:cMLKMpi} in the case of $d=4,8$. Additionally, we sample $4000$ Monte Carlo samples for testing.
Here, we only consider the inference procedure in the case of $d=4,8$ as it satisfies $p<n'$. We point out that the inference method may fail when $p>n'$. This issue arises from the non-invertibility of the $p\times p$ matrix $\mathbf{F}^{\top}\mathbf{F}$ in the covariance estimator $\widehat{\sigma}_y^2(x_{i})$ discussed in Section \ref{sec:varmlkm}. In such situations, we suggest using traditional conformal inference methods \cite[see,][]{lei2018distribution}.

Table \ref{ta:multi} summarizes the results. The MLKM outperforms the RF in terms of estimation accuracy. Regarding computational time, KRR is the most demanding, while both RF and MLKM significantly reduce time.  
Additionally, KRR requires storing a matrix of size $n\times n=4000^2$, and RF needs to store a  matrix of size $n\times D = 4000\times 500$.
In contrast, MLKM is space-efficient and only requires storing $[(d\times 256 + 256\times 16 + 16 \times 1)\times 2] + [256+16]=74032$ in space when $d=128$, where the first part represents the space for parameters and gradients, and the second part represents the space for random features.
This observation aligns with the discussions in Section \ref{sec:compcomplexity} and highlights the efficiency of  MLKM. 

\begin{table}[ht]
\caption{Comparison of KRR, RF, MLKM, and RKM in Example \ref{eg:additive1}. We include the length and coverage probability of the 95\% confidence bands. The time measure for MLKM is based on per-epoch.}
\centering
\begin{tabular}{cccccc}
 \hline
  & & KRR & RF & MLKM & RKM \\
    \hline
    $d=4$ & Testing MSE & $1.472$ & $1.778$ & $\mathbf{1.207}$ & $\mathbf{1.196}$\\
    &Time &$2.765$s & $0.211$s  & $0.031$s &$0.032$s\\
    &Confidence Band & & &  $4.192 (\mathbf{94.92\%})$ & $4.351 (\mathbf{95.33\%})$\\
    \hline
    $d=8$&  Testing MSE & $2.162$ & $2.666$ & $\mathbf{1.545}$ & $\mathbf{1.506}$\\
    &Time &$2.307$s & $0.243$s  & $0.035$s &$0.034$s\\
    &Confidence Band & & &  $4.776 (\mathbf{94.17\%})$ & $4.759 (\mathbf{94.78\%})$\\
    \hline
    $d=16$ & Testing MSE & $3.802$& $5.269$ & $\mathbf{2.548}$ & $\mathbf{2.498}$\\
    &Time &$2.551$s & $0.306$s  & $0.043$s &$0.047$s\\
    \hline
    $d=32$&  Testing MSE & $5.119 $ & $8.384 $ & $\mathbf{7.035} $ & $\mathbf{7.998}$\\
    &Time &$2.566$s & $0.333$s  & $0.048$s &$0.051$s\\
    \hline
    $d=64$& Testing MSE & $ 8.340$ & $13.977$ & $14.996$ & $14.101$\\ 
     &Time &$2.875$s & $0.291$s  & $0.056$s &$0.066$s\\
    \hline
    $d=128$&  Testing MSE & $14.899$ & $18.066$ & $\mathbf{17.009}$ & $\mathbf{16.684}$\\
     &Time &$2.709$s & $0.437$s  & $0.057$s &$0.061$s\\
    \hline
\end{tabular}
    \label{ta:multi}
\end{table}

Figure \ref{fig:multiadd} shows the 95\% confidence bands for both MLKM and RKM with $d=8$. Here we vary the values of $x_{j}$ while holding other variables $x_{-j}=1/2$, where $x_{-j}$ denotes all covariates of $X$ other than $x_{j}$. This is demonstrated through five subplots for $j\in\{1,\cdots,5\}$. Additionally, Table \ref{ta:multi} shows the average length and the empirical coverage probability of these confidence bands at the  95\% significance level. Both MLKM and RKM have coverage probabilities close to 95\%, which indicates that our conformal prediction-based method effectively generates confidence bands with finite-sample guarantees.

\begin{figure}[ht]
	\centering
	\begin{minipage}[t]{0.98\textwidth}
		\centering
		\includegraphics[width=\textwidth]{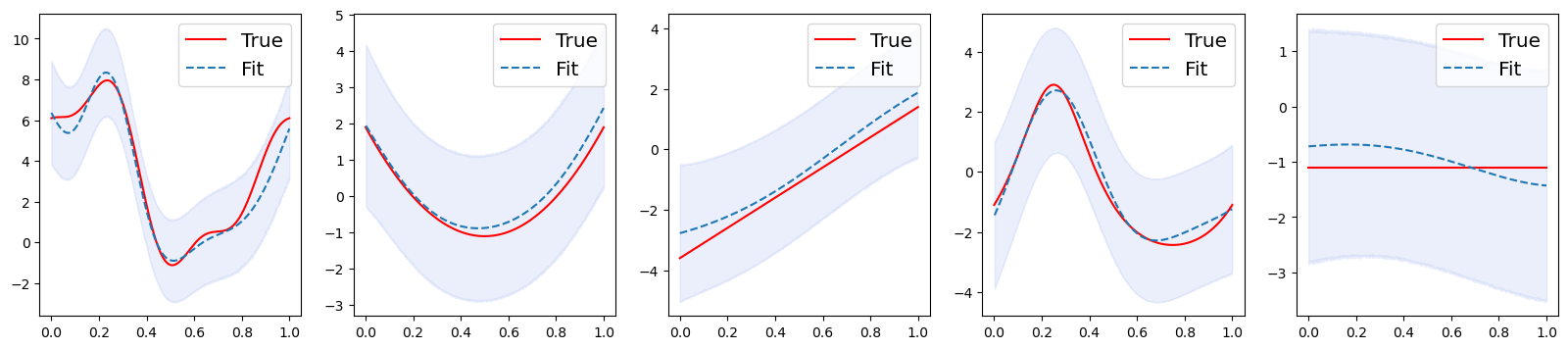}
	\end{minipage}
    \begin{minipage}[t]{0.98\textwidth}
		\centering
		\includegraphics[width=\textwidth]{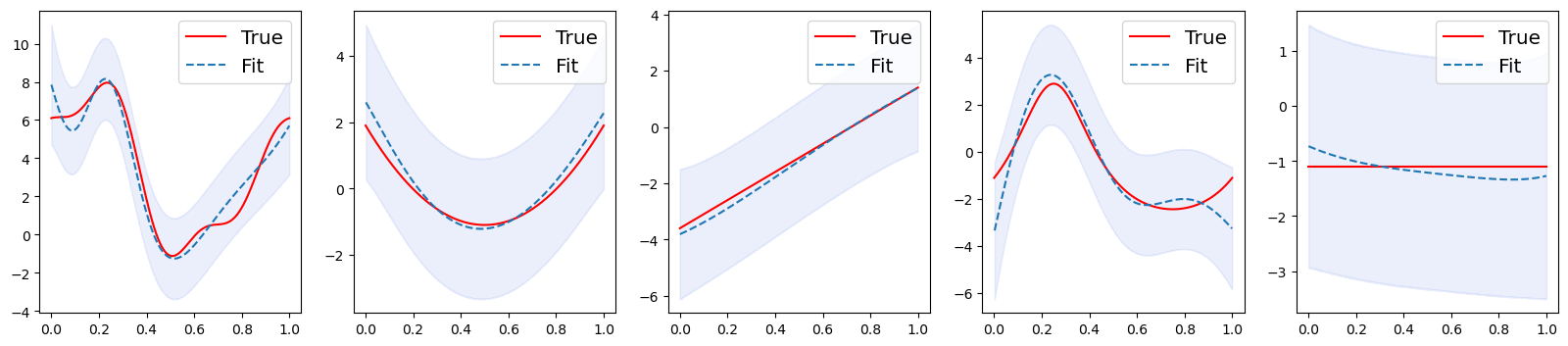}
	\end{minipage}
    \caption{Confidence bands of MLKM (upper) and RKM (lower) in Example \ref{eg:additive1}. The $j$-th subplot from left to right shows the true function $f^*(x)$ versus $x_j$, the estimated function $\widehat{f}(x)$ versus $x_j$ with the 95\% upper and lower confidence bounds. } 
	\label{fig:multiadd}
\end{figure}
\end{example}

\begin{example}
\label{eg:interactaion1}
We consider a multivariate model with interactions $f^*(x_i)=a_1f_1(x_{i1})+\sum_{j=2}^4a_j(x_{i1})f_j(x_{ij})$, 
where 
\begin{align*}
    &f_1(x_1)=-2\sin(2\pi x_1),  f_2(x_2)=x_2^2-1/3,  f_3(x_3)=x_3-1/2,  f_4(x_4)=e^{x_4}+e^{-1}-1,\\
    &a_1=1, \quad a_2(x_1)=\sqrt{\frac{2}{\pi}}e^{-\frac{(x_1-1)^2}{2}}, \quad a_3(x_1)=3\cos(2\pi x_1), \quad a_4(x_1)=4.
\end{align*}
This model is considered by \cite{ravikumar2009sparse, Lu2019Kernel}. 
Let $X_1,\cdots,X_d$ follow i.i.d. Uniform$[0,1]$. We generate data $x_{1j},\ldots,x_{nj}$ that are i.i.d. copies of $X_j$. The noise $\epsilon_i$ in \eqref{eqn:trueregression} follows $N(0,1)$. Let the dimension $d\in\{4,8\}$ and the sample size $n=4000$. Additionally, we sample $4000$ Monte Carlo samples for testing.
We compare four models as Example \ref{eg:additive1}: (i) KRR, (ii) RF, (iii) MLKM, and (iv) RKM.

Table \ref{ta:ATLA} summarizes the comparison results.
Figure \ref{fig:multipair} shows the confidence bands for both MLKM and RKM at the 95\% significance level. It is seen that our proposed methods outperform alternative methods in terms of estimation accuracy and computational complexity. 

\begin{table}[t!]
\caption{Comparison of KRR, RF, MLKM, and RKM in Example \ref{eg:interactaion1}. We include the length and coverage probability of the 95\% confidence bands. The time measure for MLKM is based on per-epoch.}
 \centering
	\begin{tabular}{cccccc}
    \hline
     & & KRR & RF & MLKM & RKM \\
    \hline 
    $d=4$ & Testing MSE & $1.066$ & $1.076$& $1.074$ &$\mathbf{1.064}$\\
    &Time & $2.711$s &$0.225$s & $0.028$s &$0.034$s\\
    &Confidence Band & & &  $4.092 (\mathbf{95.24\%})$ & $4.077 (\mathbf{95.37\%})$\\
    \hline
    $d=8$ & Testing MSE & $1.265$& $1.240$ & $\mathbf{1.207}$ & $\mathbf{1.196}$\\
    &Time & $2.451$s &$0.244$s & $0.036$s &$0.036$s\\
    &Confidence Band & & &  $ 4.298 (\mathbf{94.50\%})$ & $4.382 (\mathbf{95.13\%})$\\
    \hline $d=16$ & Testing MSE & $1.837$& $2.245$ & $\mathbf{1.940}$ & $\mathbf{1.747}$\\
    &Time &$2.618$s & $0.248$s  & $0.041$s &$0.045$s\\
    \hline $d=32$&  Testing MSE & $3.803 $ & $3.457$ & $4.426$ & $3.802$\\
    &Time &$3.165$s & $0.344$s  & $0.040$s &$0.060$s\\
    \hline $d=64$& Testing MSE & $15.554$ & $7.263$ & $7.415$ & $\mathbf{7.196}$\\ 
     &Time &$2.851$s & $0.308$s  & $0.048$s &$0.053$s\\
    \hline
    $d=128$&  Testing MSE & $148.895$ & $8.164$ & $\mathbf{8.131}$ & $\mathbf{7.537}$\\
     &Time &$2.591$s & $0.366$s  & $0.047$s &$0.047$s\\
    \hline
	\end{tabular}
    \label{ta:ATLA}
\end{table}

\begin{figure}[ht]
	\centering
    \begin{minipage}[t]{0.98\textwidth}
		\centering
		\includegraphics[width=\textwidth]{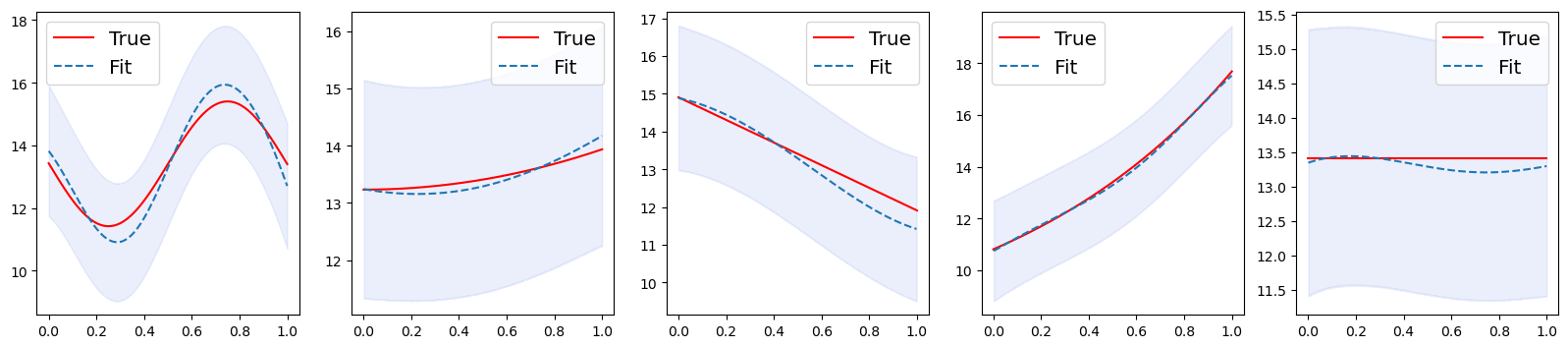}
	\end{minipage}
    \begin{minipage}[t]{0.98\textwidth}
		\centering
		\includegraphics[width=\textwidth]{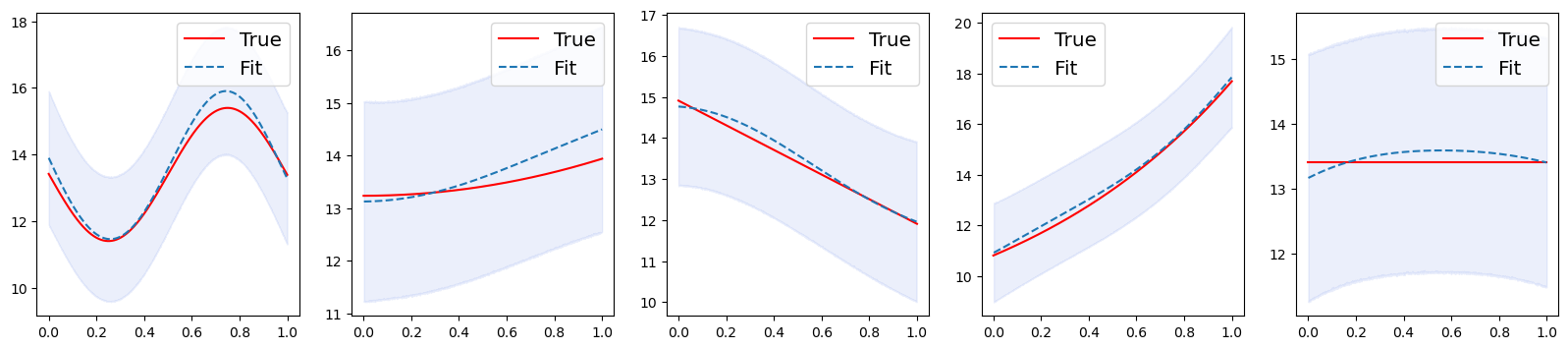}
	\end{minipage}
	\caption{Confidence bands of MLKM (upper) and RKM (lower) in Example \ref{eg:interactaion1} with $d=8$. The $j$-th subplot from left to right shows the true function $f^*(x)$ versus $x_j$, the estimated function $\widehat{f}(x)$ versus $x_j$ with the 95\% upper and lower confidence bounds. }
	\label{fig:multipair}
\end{figure}
\end{example}

\begin{example}
\label{eg:tripoly}
We consider another model $f^*(x_i)=\sum_{j=1}^{d}f_j(x_{ij})$, where the component functions are chosen from either the trigonometric polynomial or the sin-ratio functions: 
\begin{equation*}
\begin{aligned}
&\text{trigonometric polynomial:}\quad &&f_j(x_j)=u_{j,1}\sin{x_j}+u_{j,2}\cos{x_j}+u_{j,3}\sin^2{x_j};\\
&\text{sin-ratio:}\quad &&f_j(x_j)=\sin{(c_{j,1}x_j)}/[2-\sin{(c_{j,2}x_j)}].
\end{aligned}
\end{equation*}
Here $u_{j,1},u_{j,2},u_{j,3}$ and $c_{j,1},c_{j,2}$ follow i.i.d. Uniform$[1,2]$.
We consider three types of $f^*$: (i) $f_j$'s are trigonometric polynomials for $j=1,\ldots,d$; (ii) $f_j$'s are sin-ratios  for $j=1,\ldots,d$; (iii) $f_1,\ldots,f_{d/2}$ are trigonometric polynomials and $f_{d/2+1},\ldots,f_{d}$ are sin-ratios.
This model is considered by \cite{dai2022kernel}.
We generate i.i.d. copies $(x_{i1},x_{i2},\cdots,x_{id})$ from two distributions: a uniform $[-2,2]^{d}$ distribution and a multivariate normal distribution with mean zero and covariance $\Sigma_{jk}=0.5^{|j-k|}$, where $i=1,\ldots,n$.
Let the dimension $d=10$ and the sample size $n=500$. Additionally, we sample $1000$ Monte Carlo samples for testing. We compare four models: (i) a four-layer neural network (NN), (ii) residual neural network (ResNet) \citep{He2015DeepRL}, (iii) MLKM, and (iv) RKM. These models share the same layer structures (10-64-32-16-8-1). For MLKM and RKM, we select random features that correspond to Gaussian features with scale parameters $\gamma=0.25,0.5,1,2$ in each layer.

\begin{table}[ht]
\caption{Comparison of NN, ResNet, MLKM, and RKM in Example \ref{eg:tripoly}. }
	\centering
	\begin{tabular}{ccccccc}
 \hline
    & & & NN & ResNet & MLKM & RKM \\
    \hline
    (i) tri poly & uniform &testing MSE & $2.349$ & $2.030$& $\mathbf{1.939}$ & $\mathbf{1.942}$\\
    & &Time & $11.922$s &$14.899$s & $13.434$s &$18.982$s\\
     & normal &testing MSE & $3.543$ & $4.149$& $\mathbf{2.675}$ & $\mathbf{1.856}$\\
    & &Time & $11.269$s &$16.587$s &$15.002$s &$15.317$s\\
    \hline
    (ii) sin-ratio &uniform & testing MSE & $0.794$ & $0.809$& $\mathbf{0.792}$ & $\mathbf{0.761}$\\
    & &Time &$10.189$s & $14.444$s  & $15.220$s & $16.619$s\\
     & normal &testing MSE & $1.174$ & $1.334$& $\mathbf{0.923}$ & $\mathbf{1.017}$\\
    & &Time & $11.444$s &$15.134$s & $15.198$s & $16.861$s\\
    \hline
    (iii) mix  &uniform & testing MSE & $1.300$ & $1.906$& $\mathbf{1.225}$ & $1.308$\\
    & &Time &$10.992$s & $14.884$s  & $13.477$s &$15.933$s\\
     & normal &testing MSE & $2.140$ & $1.827$& $\mathbf{1.487}$ & $\mathbf{1.500}$\\
    & &Time & $11.430$s &$16.221$s &$14.002$s &$17.613$s \\
    \hline
	\end{tabular}
    \label{ta:tripoly}
\end{table}

\begin{figure}[ht]
    \centering
	\begin{minipage}[t]{0.45\textwidth}
		\centering
		\includegraphics[width=\textwidth]{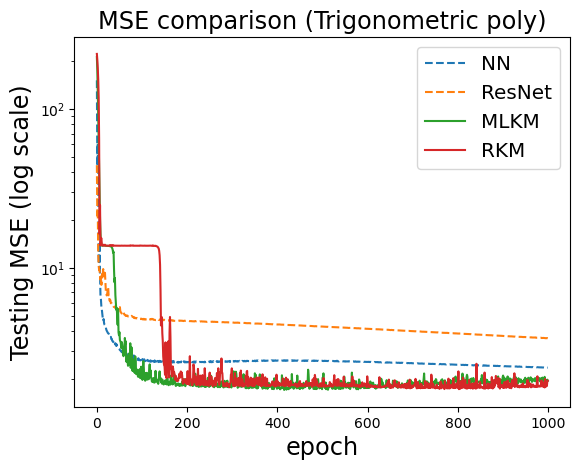}
	\end{minipage}
    \begin{minipage}[t]{0.45\textwidth}
		\centering
		\includegraphics[width=\textwidth]{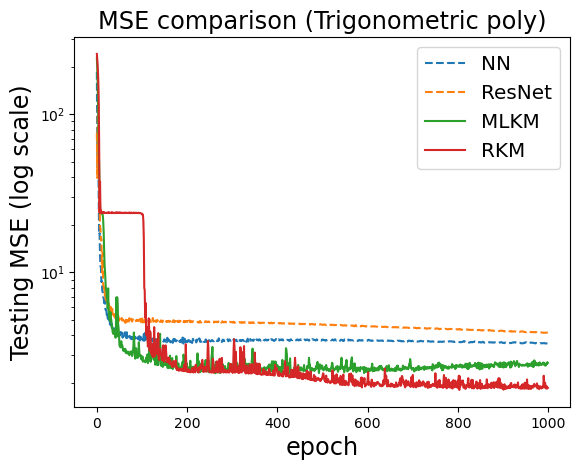}
	\end{minipage}
    \begin{minipage}[t]{0.45\textwidth}
		\centering
		\includegraphics[width=\textwidth]{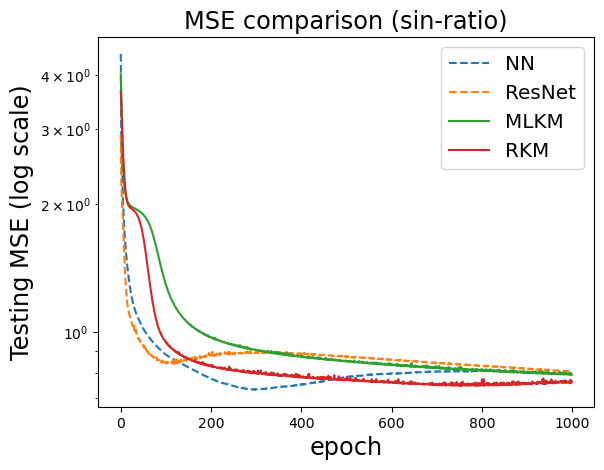}
	\end{minipage}
    \begin{minipage}[t]{0.45\textwidth}
		\centering
		\includegraphics[width=\textwidth]{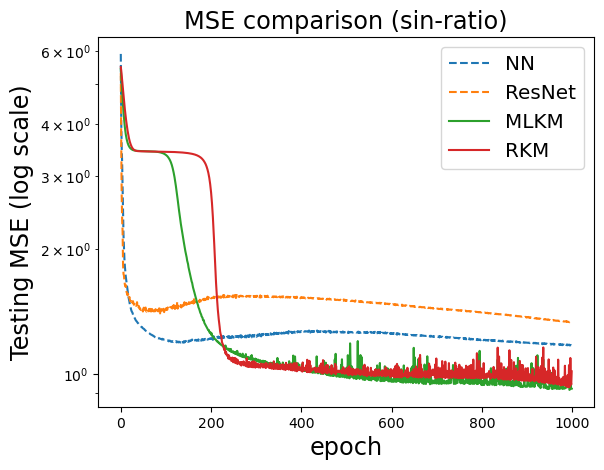}
	\end{minipage}
    \begin{minipage}[t]{0.45\textwidth}
		\centering
		\includegraphics[width=\textwidth]{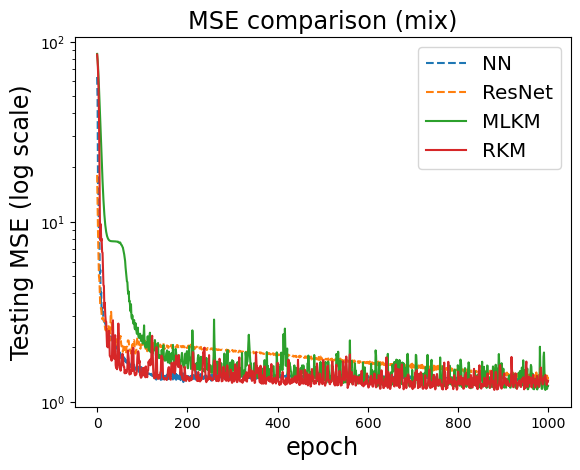}
	\end{minipage}
    \begin{minipage}[t]{0.45\textwidth}
		\centering
		\includegraphics[width=\textwidth]{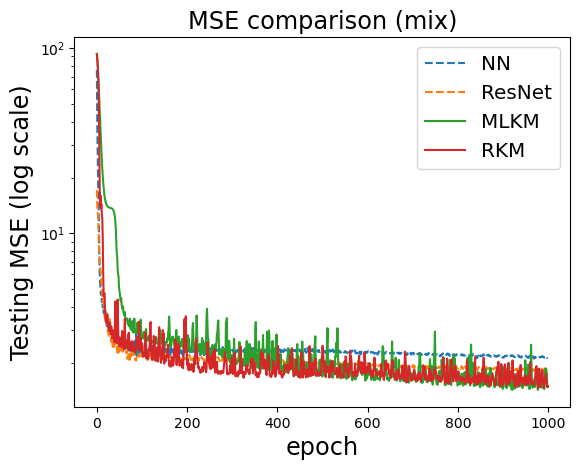}
	\end{minipage}
    \caption{Comparison of testing MSE in Example \ref{eg:tripoly}. The left and right plots correspond to covariates with uniform and normal distributions, respectively.}
    \label{fig:tripoly}
\end{figure}

Table \ref{ta:tripoly} summarizes the results. The MLKM and RKM consistently outperform the NN-based methods in terms of estimation accuracy in most cases. 
The results of RKM also demonstrate the advantage of incorporating residual learning into MLKM as discussed in Section \ref{sec:extensionMLKM}. Moreover, our MLKM method integrates a multi-scaled structure within the multi-layered architecture without introducing additional time complexity. Figure \ref{fig:tripoly} shows the convergence performance in terms of testing MSEs during model fitting. It is seen that our MLKM converges fast as illustrated in Theorem \ref{nonconvex0}. 

\end{example}

\begin{example}\label{eg:expo}
We consider the model $f^*(x_i)=\sum_{j=1}^{d}f_j(x_{ij})$, where the component functions are chosen as: 
\begin{equation*}
    f_j(x_j)=x_{j}+2 e^{-16x_{j}^2}
\end{equation*}
This model is considered by \cite{braun2005kernel}.
We generate i.i.d. copies $(x_{i1},x_{i2},\cdots,x_{id})$ from a uniform $[0,1]^{d}$ distribution, where $i=1,\ldots,n$.
Let the dimension $d \in\{1, 2, 5\}$ and the sample size $n=1000$. Additionally, we sample $1000$ Monte Carlo samples for testing. We compare four models: (i) a four-layer neural network (NN), (ii) residual neural network (ResNet) \citep{He2015DeepRL}, (iii) MLKM, and (iv) RKM. These models share the same layer structures ($d$-10-10-10-10-1). For MLKM and RKM, we select random features that correspond to Gaussian features with scale parameters $\gamma=0.5,1,2,4$ in each layer.
Table \ref{ta:d-} summarizes the results. It is seen that, in most cases, the MLKM yields estimation accuracy comparable to NN-based methods. 
\begin{table}[ht]
\caption{Comparison of NN, ResNet, MLKM, and RKM in Example \ref{eg:expo} in terms of testing MSE.}
	\centering
	\begin{tabular}{ccccc}
 \hline
    & NN & ResNet & MLKM & RKM \\
    \hline
    $d=1$& $7.80\times 10^{-5}$	&$4.00\times 10^{-5}$	&$\mathbf{1.10\times 10^{-5}}$ &$\mathbf{3.10\times 10^{-5}}$\\
    $d=2$& $7.38\times 10^{-4}$ &$4.07\times 10^{-4}$ &	$\mathbf{2.77\times 10^{-4}}$	& $\mathbf{3.86\times 10^{-4}}$\\
    $d=5$&	$1.29\times 10^{-2}$	&$5.95\times 10^{-3}$	&$3.59\times 10^{-2}$ &$\mathbf{5.51\times 10^{-3}}$\\
    \hline
	\end{tabular}
    \label{ta:d-}
\end{table}

To study the estimation performance for larger input dimensions, we consider the model $f^*(x_i)=\sum_{j=1}^{10}f_j(x_{ij})$, where the component functions are defined as above. Here, the number of active parameters 10 is fixed and is smaller than the input dimension $d$. We generate i.i.d. copies $(x_{i1},x_{i2},\cdots,x_{id})$ from a uniform $[0,1]^{d}$ distribution, where $i=1,\ldots,n$. Let the dimension $d \in\{100,1000\}$ and the sample size $n=1000$. Additionally, we sample $1000$ Monte Carlo samples for testing. We compare four models as above.
Table \ref{ta:d+} summarizes the results. The results indicate that all methods suffer from the curse of dimensionality. However, our proposed MLKM methods are consistently more adaptive with better estimation accuracy for larger input dimensions.

\begin{table}[ht]
\caption{Comparison of NN, ResNet, MLKM, and RKM in Example \ref{eg:expo} in terms of testing MSE.}
\centering
\begin{tabular}{ccccc}
\hline
    & NN & ResNet & MLKM & RKM \\
    \hline
    $d=100$ &$3.417$& $3.148$ & $\mathbf{2.881}$& $\mathbf{3.041}$\\
    $d=1000$ &	$2.971$ & $2.818$& $2.896$& $\mathbf{2.694}$ \\
    \hline
\end{tabular}
    \label{ta:d+}
\end{table}

\end{example}

\section{Real Data Examples}\label{sec:real}
We apply the proposed methods to two types of real data analysis: a temperature forecasting data analysis, and an audio feature data analysis. 
We aim to test the performance of our methods in prediction and inferential analysis under real applications.

\subsection{Temperature forecasting data analysis}
We first consider an application for indoor temperature forecasting, using the dataset SML2010 Dataset at UCI Machine Learning repository \cite{misc_sml2010_274}. The data were captured from a solar domotic house that participated in the Solar Decathlon Europe 2010, a world competition for energy efficiency. It contains $n=2764$ training samples and $1373$ testing samples, where for each sample, there are several environmental features collected from a monitor system mounted in an intelligent house. We use $d=15$ attributes as covariates, including two carbon dioxide features, two relative humidity features, two lighting features, precipitation, sun dusk, wind speed, three sunlight features, sun irradiance, outdoor temperature, and outdoor relative humidity. The response variable $y$ is set to be the indoor temperature. Our goal is to predict the indoor temperature using the internal and external environmental features, which is an important task in agricultural production and thermal energy optimization \cite{jiang2018data,mancipe2022prediction}.

We first normalize the covariates onto $[0,1]$ and compare five models: (i) random feature (RF) method \cite{2007Random}, (ii) a two-layer neural network (NN), (iii) residual neural network (ResNet) \citep{He2015DeepRL}, (iv) the proposed MLKM in Algorithm \ref{alg0}, and (v) the proposed residual-kernel machine (RKM) in \eqref{fun2}. The latter four models share the same layer structures ($d$-100-50-1). For RF, we select $D=500$ Gaussian features with scale parameter $\gamma=0.4$. For MLKM and RKM, we select Gaussian random features with scale parameters $\gamma=0.1,0.4$ in each layer. We randomly split the data into two subsets with sizes $\{n',m\}$, where $n'=m=n/2$. We use $n'$ data to train the MLKM machine and use $m$ data to construct the conformal MLKM prediction interval in \eqref{eqn:cMLKMpi}.

Table \ref{ta:sml} summarizes the results. It shows that our proposed MLKM method achieves better estimation accuracy compared to NN-based methods. Besides, while the neural networks exhibit smaller training errors, they suffer from over-fitting issues compared to the MLKM. Moreover, MLKM outperforms the single-layer random feature method, demonstrating the advantages of multiple layers and multiple scales. Additionally, Table \ref{ta:sml} shows the average length and the empirical coverage probability of the confidence bands at the  95\% significance level. The true temperature lies within a temperature error of $3.804^\circ$C for MLKM and $3.116^\circ$C for RKM predictions, with a probability near $95\%$. Our conformal prediction-based method has desirable finite-sample guarantees.

\begin{table}[ht]
\caption{Comparison of RF, NN, ResNet, MLKM, and RKM in SML2010 dataset. We include the length and coverage probability of the 95\% confidence bands.}
	\centering
    \begin{tabular}{cccccc}
    \hline
     & RF & NN &ResNet  & MLKM & RKM  \\
    \hline
    Training MSE & $0.122$ &  $0.373$ & $0.240$ & $0.386$ &  $0.497$  \\
    Testing MSE  & $3.958$& $1.693$ & $1.554$  & $\mathbf{1.518}$ & $\mathbf{1.347}$ \\
    Confidence Band & & & & $7.608(89.38\%)$ & $6.233 (91.54\%)$ \\
    \hline
	\end{tabular}
    \label{ta:sml}
\end{table}

\subsection{Audio feature data analysis}\label{experiment}
We consider another application with audio feature data by using the MillionSongs dataset \cite{misc_yearpredictionmsd_203}. This dataset contains $463715$ training samples and $51630$ test samples. For each sample, there are $d=90$ attributes about the audio features of a song, along with a response variable $y$ of its release year ranging from 1922 to 2011. The attributes are metadata extracted from the timbre features of different music segments, which represent the average and covariance of the timbre vectors for each song. Our goal is to predict the year a song was composed using its musical features, which is important in music information retrieval (MIR) and music recommendation \cite{bertin2011million}. 
We normalize the covariates onto $[0,1]$ and compare four models: (i) a two-layer neural network (NN), (ii) residual neural network (ResNet) \citep{He2015DeepRL}, (iii) the proposed MLKM in Algorithm \ref{alg0}, and (iv) the proposed residual-kernel machine (RKM) in \eqref{fun2}. These models share the same layer structures ($d$-32-8-1).  For MLKM and RKM, we select random features corresponding to the Cauchy random features and Gaussian random features with scale parameters $\gamma=0.01,0.1$, respectively, in two layers. We randomly select two subsets in the training set with sizes  $\{n',m\}$, where $n'=700,m=500$. We use $n'$ data to construct the MLKM machine and use $m$ data to construct the conformal MLKM prediction interval. Additionally, we sample 300 data for testing. 

Firstly, we investigate the advantages of incorporating the residual learning structure as well as the cross-fitting technique in Algorithm \ref{alg0}. 
Table \ref{ta1} summarizes the results. It is seen that our proposed MLKM method outperforms NN-based methods. Additionally, incorporating residual blocks and the cross-fitting technique further improves the estimation performance. Meanwhile, the running time does not increase significantly compared to neural networks.
\begin{table}[ht]
\caption{Comparison of NN, ResNet, MLKM, and RKM in MillionSongs dataset. We use ``MLKM" to denote the case without cross-fitting and ``MLKM $+$" to denote the case with cross-fitting.}
	\centering
     \begin{tabular}{cccccc}
         \hline
         & NN &ResNet & MLKM & RKM & MLKM $+$    \\
        \hline
        Training MSE & $ 29.858  $ & $ 23.882  $  & $ 99.232$ & $ 112.295 $ & $ 134.137$\\
        Testing MSE  & $ 980.874 $ & $ 1486.201$ & $\mathbf{236.850}$ & $\mathbf{179.474}$ & $ \mathbf{173.256} $ \\
        \hline
    \end{tabular}
    \label{ta1}
\end{table}

Secondly, we study the statistical inference performance. We compare three inference methods: (i) Bootstrap, (ii) the split-conformal prediction interval \cite{lei2018distribution}, and (iii) our conformal MLKM prediction interval in \eqref{eqn:cMLKMpi}.
We conduct the Bootstrap method as a benchmark. Table \ref{ta3} shows the average length and the empirical coverage probability of the confidence bands at the 95\% significance level for different methods. 
It is seen that our model-based conformal MLKM bands yield better coverage probability than other methods.

\begin{table}[ht]
\caption{Comparison of NN's, ResNet's, MLKM's, and RKM's confidence bands for different methods in MillionSongs dataset. We include the length and coverage probability of the 95\% confidence bands.}
	\centering
	\begin{tabular}{ccccc}
 \hline
     & NN & ResNet & MLKM  & RKM\\
    \hline
    Bootstrap Band & $ 107.71 (91.33\%) $ & $108.98 (91.00\%)$ & $46.51 (81.00\%) $ & $20.02 (60.67\%)$\\
    Split-Conformal Band  &  & & $ 47.64 (86.33\%) $ & $ 35.21 (82.00\%)$\\
    Conformal MLKM Band  & &  & $52.55 (\mathbf{90.62\%})  $ & $41.15 (83.29\%) $\\
    \hline
	\end{tabular}
    \label{ta3}
\end{table}

Finally, we carry out a large-scale experiment. We compare four models as above: three-layer NN, ResNet, MLKM, and RKM. Now these models share the same layer structures (d-256-128-64-1). Table \ref{ta4} summarizes the results. It is seen that our proposed MLKM models achieve better estimation performance and are still efficient in large-scale learning settings.
\begin{table}[ht]
\caption{Comparison of NN, ResNet, MLKM, and RKM in large-scale MillionSongs dataset.}
	\centering
	\begin{tabular}{ccccc}
 \hline
    & NN &ResNet & MLKM& RKM\\
    \hline
    Test MSE & $376.083$ & $376.102$ & $\mathbf{347.164}$ & $ \mathbf{351.168}$\\
    Total Time  & $212.476$s & $292.899$s & $169.651$s  & $270.103$s \\
    \hline
	\end{tabular}
    \label{ta4}
\end{table}

\section{Conclusion}\label{sec:summary}
In this paper, we introduced the Multi-Layer Kernel Machine (MLKM), a novel framework that combines random feature mapping with multiple kernel learning in a novel multi-layer architecture. We show that MLKM achieves the minimax optimal rate in estimation complex functions, and enjoys a significant reduction in computational cost. 
We demonstrate that MLKM achieves the minimax optimal rate in estimating complex functions and offers a considerable reduction in computational costs. Additionally, we have developed a robust uncertainty quantification approach that provides a valid coverage guarantee for MLKM. 
The code for reproducing all numerical results is available at \url{https://github.com/ZZZhyEva/Multi-Layer-Kernel-Machine}. We also offer a Python package for implementing MLKM at \url{https://pypi.org/project/Multi-Layer-Kernel-Machine/}.

\clearpage
\baselineskip=17pt
\bibliographystyle{apa}
\bibliography{ref}

\begin{thebibliography}{}

\bibitem[\protect\astroncite{Al-Shedivat et~al.}{2017}]{2016Learning}
Al-Shedivat, M., Wilson, A.~G., Saatchi, Y., Hu, Z., and Xing, E.~P. (2017).
\newblock Learning scalable deep kernels with recurrent structure.
\newblock {\em Journal of Machine Learning Research}, 18(1):2850--2886.

\bibitem[\protect\astroncite{Angelopoulos and Bates}{2023}]{2023A}
Angelopoulos, A.~N. and Bates, S. (2023).
\newblock Conformal prediction: A gentle introduction.
\newblock {\em Found. Trends Mach. Learn.}, 16(4):494–591.

\bibitem[\protect\astroncite{Aronszajn}{1950}]{aronszajn1950theory}
Aronszajn, N. (1950).
\newblock Theory of reproducing kernels.
\newblock {\em Transactions of the American Mathematical Society},
  68(3):337--404.

\bibitem[\protect\astroncite{Bach}{2008}]{bach2008consistency}
Bach, F.~R. (2008).
\newblock Consistency of the group lasso and multiple kernel learning.
\newblock {\em Journal of Machine Learning Research}, 9(6).

\bibitem[\protect\astroncite{Barber et~al.}{2023}]{barber2023conformal}
Barber, R.~F., Candes, E.~J., Ramdas, A., and Tibshirani, R.~J. (2023).
\newblock Conformal prediction beyond exchangeability.
\newblock {\em The Annals of Statistics}, 51(2):816--845.

\bibitem[\protect\astroncite{Bertin-Mahieux}{2011}]{misc_yearpredictionmsd_203}
Bertin-Mahieux, T. (2011).
\newblock {YearPredictionMSD}.
\newblock UCI Machine Learning Repository.
\newblock {DOI}: https://doi.org/10.24432/C50K61.

\bibitem[\protect\astroncite{Bertin-Mahieux et~al.}{2011}]{bertin2011million}
Bertin-Mahieux, T., Ellis, D.~P., Whitman, B., and Lamere, P. (2011).
\newblock The million song dataset.

\bibitem[\protect\astroncite{Bohn et~al.}{2019}]{bohn2017representer}
Bohn, B., Rieger, C., and Griebel, M. (2019).
\newblock A representer theorem for deep kernel learning.
\newblock {\em Journal of Machine Learning Research}, 20(1):2302--2333.

\bibitem[\protect\astroncite{Bottou et~al.}{2018}]{bottou2018optimization}
Bottou, L., Curtis, F.~E., and Nocedal, J. (2018).
\newblock Optimization methods for large-scale machine learning.
\newblock {\em SIAM Review}, 60(2):223--311.

\bibitem[\protect\astroncite{Braun and Huang}{2005}]{braun2005kernel}
Braun, W.~J. and Huang, L.-S. (2005).
\newblock Kernel spline regression.
\newblock {\em Canadian Journal of Statistics}, 33(2):259--278.

\bibitem[\protect\astroncite{Chernozhukov et~al.}{2018}]{2017Double}
Chernozhukov, V., Chetverikov, D., Demirer, M., Duflo, E., Hansen, C., Newey,
  W., and Robins, J. (2018).
\newblock Double/debiased machine learning for treatment and structural
  parameters.
\newblock {\em The Econometrics Journal}, 21(1):C1--C68.

\bibitem[\protect\astroncite{Cho and Saul}{2009}]{cho2009kernel}
Cho, Y. and Saul, L. (2009).
\newblock Kernel methods for deep learning.
\newblock {\em Advances in Neural Information Processing Systems}, 22.

\bibitem[\protect\astroncite{Chryssolouris and Lee}{1996}]{1996Confidence}
Chryssolouris, G. and Lee, M. (1996).
\newblock Confidence interval prediction for neural network models.
\newblock {\em IEEE Transactions on Neural Networks}, 7(1):229.

\bibitem[\protect\astroncite{Dai and Li}{2022}]{Dai2021OrthogonalizedKD}
Dai, X. and Li, L. (2022).
\newblock Orthogonalized kernel debiased machine learning for multimodal data
  analysis.
\newblock {\em Journal of the American Statistical Association}, pages 1--15.

\bibitem[\protect\astroncite{Dai et~al.}{2022}]{dai2022kernel}
Dai, X., Lyu, X., and Li, L. (2022).
\newblock Kernel knockoffs selection for nonparametric additive models.
\newblock {\em Journal of the American Statistical Association}, pages 1--13.

\bibitem[\protect\astroncite{Damianou and Lawrence}{2013}]{damianou2013deep}
Damianou, A. and Lawrence, N.~D. (2013).
\newblock Deep gaussian processes.
\newblock In {\em Artificial Intelligence and Statistics}, pages 207--215.
  PMLR.

\bibitem[\protect\astroncite{Daubechies}{1992}]{daubechies1992ten}
Daubechies, I. (1992).
\newblock {\em Ten Lectures on Wavelets}.
\newblock SIAM.

\bibitem[\protect\astroncite{Dunlop et~al.}{2018}]{2017How}
Dunlop, M.~M., Girolami, M.~A., Stuart, A.~M., and Teckentrup, A.~L. (2018).
\newblock How deep are deep gaussian processes?
\newblock {\em Journal of Machine Learning Research}, 19(54):1--46.

\bibitem[\protect\astroncite{Gale et~al.}{2019}]{Gale2019TheSO}
Gale, T., Elsen, E., and Hooker, S. (2019).
\newblock The state of sparsity in deep neural networks.
\newblock {\em arXiv preprint arXiv:1902.09574}.

\bibitem[\protect\astroncite{Geer}{2000}]{geer2000empirical}
Geer, S.~A. (2000).
\newblock {\em Empirical Processes in M-estimation}, volume~6.
\newblock Cambridge University Press.

\bibitem[\protect\astroncite{G{\"o}nen and
  Alpayd{\i}n}{2011}]{gonen2011multiple}
G{\"o}nen, M. and Alpayd{\i}n, E. (2011).
\newblock Multiple kernel learning algorithms.
\newblock {\em The Journal of Machine Learning Research}, 12:2211--2268.

\bibitem[\protect\astroncite{Halko et~al.}{2011}]{halko2011finding}
Halko, N., Martinsson, P.-G., and Tropp, J.~A. (2011).
\newblock Finding structure with randomness: Probabilistic algorithms for
  constructing approximate matrix decompositions.
\newblock {\em SIAM Review}, 53(2):217--288.

\bibitem[\protect\astroncite{Hastie et~al.}{2009}]{Hastie2008THE}
Hastie, T., Tibshirani, R., Friedman, J.~H., and Friedman, J.~H. (2009).
\newblock {\em The Elements of Statistical Learning: Data Mining, Inference,
  and Prediction}, volume~2.
\newblock Springer.

\bibitem[\protect\astroncite{He et~al.}{2016}]{He2015DeepRL}
He, K., Zhang, X., Ren, S., and Sun, J. (2016).
\newblock Deep residual learning for image recognition.
\newblock In {\em Proceedings of the IEEE conference on Computer Vision and
  Pattern Recognition}, pages 770--778.

\bibitem[\protect\astroncite{Jiang et~al.}{2018}]{jiang2018data}
Jiang, Z., Francis, J., Sahu, A.~K., Munir, S., Shelton, C., Rowe, A., and
  Berg{\'e}s, M. (2018).
\newblock Data-driven thermal model inference with armax, in smart
  environments, based on normalized mutual information.
\newblock In {\em 2018 Annual American Control Conference (ACC)}, pages
  4634--4639. IEEE.

\bibitem[\protect\astroncite{Kimeldorf and Wahba}{1971}]{KIMELDORF197182}
Kimeldorf, G. and Wahba, G. (1971).
\newblock Some results on tchebycheffian spline functions.
\newblock {\em Journal of Mathematical Analysis and Applications},
  33(1):82--95.

\bibitem[\protect\astroncite{Koltchinskii and
  Yuan}{2010}]{koltchinskii2010sparsity}
Koltchinskii, V. and Yuan, M. (2010).
\newblock Sparsity in multiple kernel learning.
\newblock {\em Annals of Statistics}, 38(6):3660--3695.

\bibitem[\protect\astroncite{Lanckriet et~al.}{2004}]{lanckriet2004learning}
Lanckriet, G.~R., Cristianini, N., Bartlett, P., Ghaoui, L.~E., and Jordan,
  M.~I. (2004).
\newblock Learning the kernel matrix with semidefinite programming.
\newblock {\em Journal of Machine Learning Research}, 5(Jan):27--72.

\bibitem[\protect\astroncite{Lei et~al.}{2018}]{lei2018distribution}
Lei, J., G’Sell, M., Rinaldo, A., Tibshirani, R.~J., and Wasserman, L.
  (2018).
\newblock Distribution-free predictive inference for regression.
\newblock {\em Journal of the American Statistical Association},
  113(523):1094--1111.

\bibitem[\protect\astroncite{Liu et~al.}{2021}]{2021Random}
Liu, F., Huang, X., Chen, Y., and Suykens, J.~A. (2021).
\newblock Random features for kernel approximation: A survey on algorithms,
  theory, and beyond.
\newblock {\em IEEE Transactions on Pattern Analysis and Machine Intelligence},
  44(10):7128--7148.

\bibitem[\protect\astroncite{Lu et~al.}{2020}]{Lu2019Kernel}
Lu, J., Kolar, M., and Liu, H. (2020).
\newblock Kernel meets sieve: post-regularization confidence bands for sparse
  additive model.
\newblock {\em Journal of the American Statistical Association},
  115(532):2084--2099.

\bibitem[\protect\astroncite{Mallat}{2016}]{mallat2016understanding}
Mallat, S. (2016).
\newblock Understanding deep convolutional networks.
\newblock {\em Philosophical Transactions of the Royal Society A: Mathematical,
  Physical and Engineering Sciences}, 374(2065):20150203.

\bibitem[\protect\astroncite{Mancipe-Castro and
  Guti{\'e}rrez-Carvajal}{2022}]{mancipe2022prediction}
Mancipe-Castro, L. and Guti{\'e}rrez-Carvajal, R. (2022).
\newblock Prediction of environment variables in precision agriculture using a
  sparse model as data fusion strategy.
\newblock {\em Information Processing in Agriculture}, 9(2):171--183.

\bibitem[\protect\astroncite{Mendelson}{2002}]{mendelson2002geometric}
Mendelson, S. (2002).
\newblock Geometric parameters of kernel machines.
\newblock In {\em International Conference on Computational Learning Theory},
  pages 29--43. Springer.

\bibitem[\protect\astroncite{Rahimi and Recht}{2007}]{2007Random}
Rahimi, A. and Recht, B. (2007).
\newblock Random features for large scale kernel machines.
\newblock In {\em Advances in Neural Information Processing Systems},
  volume~20.

\bibitem[\protect\astroncite{Raskutti et~al.}{2014}]{raskutti2014early}
Raskutti, G., Wainwright, M.~J., and Yu, B. (2014).
\newblock Early stopping and non-parametric regression: an optimal
  data-dependent stopping rule.
\newblock {\em Journal of Machine Learning Research}, 15(1):335--366.

\bibitem[\protect\astroncite{Rasmussen and
  Williams}{2006}]{rasmussen2006gaussian}
Rasmussen, C.~E. and Williams, C.~K. (2006).
\newblock {\em Gaussian Processes for Machine Learning}.
\newblock Springer.

\bibitem[\protect\astroncite{Ravikumar et~al.}{2009}]{ravikumar2009sparse}
Ravikumar, P., Lafferty, J., Liu, H., and Wasserman, L. (2009).
\newblock Sparse additive models.
\newblock {\em Journal of the Royal Statistical Society Series B: Statistical
  Methodology}, 71(5):1009--1030.

\bibitem[\protect\astroncite{Romeu-Guallart and
  Zamora-Martinez}{2014}]{misc_sml2010_274}
Romeu-Guallart, P. and Zamora-Martinez, F. (2014).
\newblock {SML2010}.
\newblock UCI Machine Learning Repository.
\newblock {DOI}: https://doi.org/10.24432/C5RS3S.

\bibitem[\protect\astroncite{Rudi and Rosasco}{2017}]{Rudi2016GeneralizationPO}
Rudi, A. and Rosasco, L. (2017).
\newblock Generalization properties of learning with random features.
\newblock In {\em Advances in Neural Information Processing Systems},
  volume~30.

\bibitem[\protect\astroncite{Rybak et~al.}{2023}]{zhou2023inference}
Rybak, J., Battey, H., and Zhou, W.-X. (2023).
\newblock On inference for the support vector machine.
\newblock {\em Preprint}, pages 1--51.

\bibitem[\protect\astroncite{Saunders et~al.}{1998}]{saunders1998ridge}
Saunders, C., Gammerman, A., and Vovk, V. (1998).
\newblock Ridge regression learning algorithm in dual variables.
\newblock In {\em International Conference on Machine Learning}.

\bibitem[\protect\astroncite{Schmidt-Hieber}{2020}]{SchmidtHieber2017NonparametricRU}
Schmidt-Hieber, J. (2020).
\newblock Nonparametric regression using deep neural networks with relu
  activation function.
\newblock {\em The Annals of Statistics}, 48(4):1875--1897.

\bibitem[\protect\astroncite{Seber and Lee}{2003}]{2012Linear}
Seber, G.~A. and Lee, A.~J. (2003).
\newblock {\em Linear Regression Analysis}, volume 330.
\newblock John Wiley \& Sons.

\bibitem[\protect\astroncite{Shekhar and Javidi}{2022}]{shekhar2020multi}
Shekhar, S. and Javidi, T. (2022).
\newblock Multi-scale zero-order optimization of smooth functions in an rkhs.
\newblock In {\em 2022 IEEE International Symposium on Information Theory},
  pages 288--293. IEEE.

\bibitem[\protect\astroncite{Vershynin}{2018}]{vershynin2018high}
Vershynin, R. (2018).
\newblock {\em High-dimensional probability: An introduction with applications
  in data science}, volume~47.
\newblock Cambridge university press.

\bibitem[\protect\astroncite{Vovk et~al.}{2005}]{vovk2005algorithmic}
Vovk, V., Gammerman, A., and Shafer, G. (2005).
\newblock {\em Algorithmic Learning in a Random World}, volume~29.
\newblock New York: Springer.

\bibitem[\protect\astroncite{Wahba}{1983}]{wahba1983bayesian}
Wahba, G. (1983).
\newblock Bayesian ``confidence intervals" for the cross-validated smoothing
  spline.
\newblock {\em Journal of the Royal Statistical Society: Series B
  (Methodological)}, 45(1):133--150.

\bibitem[\protect\astroncite{Wahba}{1990}]{wahba1990spline}
Wahba, G. (1990).
\newblock {\em Splines Models for Observational Data}.
\newblock SIAM, Philadelphia, PA.

\bibitem[\protect\astroncite{Wei et~al.}{2017}]{2019Early}
Wei, Y., Yang, F., and Wainwright, M.~J. (2017).
\newblock Early stopping for kernel boosting algorithms: a general analysis
  with localized complexities.
\newblock In {\em Advances in Neural Information Processing Systems},
  volume~30.

\bibitem[\protect\astroncite{Wilson et~al.}{2016}]{Wilson2015DeepKL}
Wilson, A.~G., Hu, Z., Salakhutdinov, R., and Xing, E.~P. (2016).
\newblock Deep kernel learning.
\newblock In {\em Artificial intelligence and statistics}, pages 370--378.
  PMLR.

\bibitem[\protect\astroncite{Yang et~al.}{2017}]{2017yangrandomized}
Yang, Y., Pilanci, M., and Wainwright, M.~J. (2017).
\newblock {Randomized sketches for kernels: Fast and optimal nonparametric
  regression}.
\newblock {\em The Annals of Statistics}, 45(3):991 -- 1023.

\bibitem[\protect\astroncite{Yin et~al.}{2019}]{yin2019rademacher}
Yin, D., Kannan, R., and Bartlett, P. (2019).
\newblock Rademacher complexity for adversarially robust generalization.
\newblock In {\em International Conference on Machine Learning}, pages
  7085--7094. PMLR.

\bibitem[\protect\astroncite{Yu et~al.}{2024}]{zhou2024estimation}
Yu, M., Wang, Y., Xie, S., and Tan, Kean Ming~Zhou, W.-X. (2024).
\newblock Estimation and inference for nonparametric expected shortfall
  regression over rkhs.
\newblock {\em Preprint}, pages 1--35.

\bibitem[\protect\astroncite{Zhang and Lin}{2006}]{zhang2006component}
Zhang, H.~H. and Lin, Y. (2006).
\newblock Component selection and smoothing for nonparametric regression in
  exponential families.
\newblock {\em Statistica Sinica}, pages 1021--1041.

\end{thebibliography}

\end{document}